\documentstyle[12pt,aasms4]{article}
\newcommand{\etal}{{\it et al.\ }}

\begin{document}

\slugcomment {Accepted for publication in the February 2000 issue of the Astronomical Journal}

\title{The Dwarf Spheroidal Companions to M31: WFPC2 Observations of 
Andromeda II\footnote{Based on observations with the NASA/ESA {\it 
Hubble Space Telescope}, obtained at the Space Telescope Science Institute,
which is operated by the Association of Universities for Research in 
Astronomy, Inc., (AURA), under NASA Contract NAS 5-26555.}
\footnote{Based, in part, on observations obtained at Kitt Peak National 
Observatory, a division of the National Optical Astronomy Observatories,
which is operated by AURA Inc.\ under cooperative agreement with the National
Science Foundation.}}

\author{G. S. Da Costa}
\affil{Research School of Astronomy \& Astrophysics, Institute of Advanced
Studies, The Australian National
University, Private Bag, Weston Creek Post Office, ACT 2611, Australia\\
Electronic mail: gdc@mso.anu.edu.au}
\authoremail{gdc@mso.anu.edu.au}

\author{T. E. Armandroff}
\affil{Kitt Peak National Observatory, National Optical Astronomy 
Observatories,\\
P.O. Box 26732, Tucson, Arizona 85726\\ Electronic mail: armand@noao.edu}
\authoremail{armand@noao.edu}


\author{Nelson Caldwell}
\affil{F. L. Whipple Observatory, Smithsonian Institution, P.O. Box 97,
Amado, Arizona 85645\\ Electronic mail: caldwell@flwo99.sao.arizona.edu}
\authoremail{caldwell@flwo99.sao.arizona.edu}

\and
 
\author{Patrick Seitzer}
\affil{Department of Astronomy, University of Michigan, Ann Arbor, Michigan
48109\\ Electronic mail: seitzer@astro.lsa.umich.edu}
\authoremail{seitzer@astro.lsa.umich.edu}

\begin{abstract}

The {\it Hubble Space Telescope} WFPC2 camera has been used to image 
Andromeda~II, a dwarf spheroidal (dSph) companion to M31.  The resulting
color-magnitude (c-m) diagrams reveal the morphology of the horizontal 
branch (HB) 
in this dwarf galaxy.  We find that like Andromeda~I, and like most of the
Galactic dSph companions, the HB morphology of And~II is predominantly red.
Unlike And~I, however, there is no evidence for a radial gradient in HB
morphology in the And~II data.
Based on a comparison with a combination of standard Galactic globular cluster
c-m diagrams scaled to reproduce the And~II mean abundance and abundance
dispersion, we interpret the observed HB morphology of And~II as indicating 
that at least 50\% of the total stellar population is younger than the 
age of the globular clusters.  This inference is
strengthened by the small number of confirmed upper-AGB carbon stars in
And~II\@.  The relatively faint luminosities (M$_{bol}$ $\approx$ --4.1) 
of these stars, 
however, suggest an age or ages nearer 6--9 Gyr, rather than 1--3 Gyr, for 
this population.  On the other hand, the existence of blue HB and RR~Lyrae 
variable stars in And~II argues for the existence of an additional
old (age $>$ 10 Gyr)
population in this dSph.  Thus And~II has had an extended epoch of star
formation like many of the Galactic dSphs.  The mean magnitude of the blue
HB in And~II suggests (m--M)$_{0}$ = 24.17 $\pm$ 0.06 and that And~II is
125 $\pm$ 60 kpc closer than M31 along the line-of-sight.  This confirms
the association of And~II with M31, rather than with M33 to which And~II lies
closer on the sky.  The true distance of And~II from the center of M31 is
between $\sim$160 and $\sim$230 kpc, comparable to the Galactocentric
distances of Fornax and of the Leo dSphs.  With the current samples of 
dSph companions,  the size of the Galaxy's and M31's dSph satellite systems
are comparable, with outer radii of order 250 kpc.  
The And~II red giant branch colors yield a mean abundance 
of $<$[Fe/H]$>$ = --1.49 $\pm$ 0.11 and a surprisingly large
internal abundance spread, which can be characterised by 
$\sigma_{int}$([Fe/H]) $\approx$ 0.36 dex.  Both these values are in good 
agreement with the recent ground-based spectroscopic study of 
C\^{o}t\'{e} \etal [1999, AJ, 118, 1645].  The And~II abundance dispersion 
found here is considerably larger than that derived for And~I from an identical 
analysis of similar data ($\sigma_{int}$([Fe/H]) = 0.21 dex).  Thus despite 
having very similar luminosities and mean metal abundances, these two M31 dSph 
companions have clearly had different chemical evolution histories.  We find
that we cannot model the abundance distribution in And~II with single
component simple chemical enrichment models.  However, we can reproduce the
form of the distribution if we assume two components, each with a simple model
abundance distribution.  The ``metal-poor'' component has 
mean abundance log($<\!z\!>$/$z_{sun}$) = --1.6, while the ``metal-rich'' one
has mean abundance log($<\!z\!>$/$z_{sun}$) = --0.95 and is outnumbered
by the metal-poor population by a ratio of $\sim$2.3 to 1.
We end by concluding that the diversity of evolutionary histories evident
among the Galactic dSph companions is now also firmly established among 
the dSph satellites of M31.  An Appendix discusses minor revisions to our
earlier And~I results that arise from the calibration and analysis techniques
adopted in this paper.  In particular, our comparisons with ground-based
photometry indicate that the zeropoint for the WFPC2 $F450W$ to $B$
transformation should be modified, by 0.055 mag, to produce fainter $B$ 
magnitudes and thus redder $B-V$ colors.

\end{abstract}


\section{Introduction}

The first Galactic dwarf spheroidal (dSph) companion galaxy, Sculptor, was
discovered in 1938 by Harlow Shapley (\markcite{HS38}Shapley 1938).  Shortly
afterwards \markcite{BH39}Baade \& Hubble (1939) reported that the Sculptor
system contained (globular) cluster-type short period variables, large
numbers of ``ordinary giants'', and that amongst the brighter Sculptor stars,
extremely blue or extremely red stars were lacking.  Thus began the study
of the stellar populations of dSph galaxies.  

By the 1990s the Galaxy was known to have nine dSph companions whose
stellar populations have been studied in considerable detail, using both
ground-based and {\it Hubble Space Telescope} (HST) observations.  
Undoubtedly, the most
unexpected result of these studies is the fascinating diversity of star
formation histories they have revealed (see, for example, recent reviews by \markcite{GD98}Da~Costa 1998 and \markcite{EG99}Grebel 1999).
The Galactic dSph galaxies have stellar populations that range from purely
or predominantly old (e.g.\ Ursa Minor) through systems such as Leo~I,
in which the bulk of the population has ages between 1 and 7 Gyr 
(\markcite{G99A}Gallart \etal 1999a, \markcite{G99B}b).  While there is some
indication of a correlation between intermediate-age population fraction
and Galactocentric distance (e.g.\ \markcite{SB94}van den Bergh 1994), 
which hints at the influence of the ``parent'' galaxy on the evolution
of the dSph satellites, 
we have at this moment in time 
little understanding of what lies behind this diversity 
of star formation histories.

One way to improve our knowledge of the origin of this diversity is to study
the dSph satellite systems around other galaxies.  The dSph companions
to M31, which after the recent discoveries of 
\markcite{TA98}Armandroff \etal (1998), \markcite{AJ99}Armandroff \etal (1999) 
and \markcite{KK99}Karachentsev \& Karachentseva (1999) now number
six in total, provide the nearest examples of this class of galaxy beyond
the Milky Way.  From the ground, studies of these M31 dSph systems are
limited to the upper $\sim$2 mags of the red giant branch.  With the WFPC2
camera aboard HST, however, photometry can be obtained for significantly 
fainter stars while at the same time the accuracy for the
brighter stars can be considerably improved.  Indeed with HST/WFPC2 
it is possible to
reach sufficiently faint that color-magnitude (c-m) diagrams can be produced
which
reveal the morphology of the horizontal branch in these systems.  When coupled
with the mean abundance and intrinsic abundance spread, the
horizontal branch morphology can then yield important information
on the star formation history.  Further, the HST/WFPC2 photometry,
because of the significantly reduced effects of image crowding, allows a much 
better determination of the giant branch intrinsic width.
This in turn can provide
important information on the chemical enrichment processes that took place.  Finally, the WFPC2 photometry can also be used to place limits 
on the age of, or even detect, the youngest main sequence population present.

In \markcite{DA96}Da~Costa \etal (1996, hereafter Paper~I) we presented WFPC2
photometry for the M31 dSph companion And~I, and showed, {\it inter alia}, 
that this dSph has
a dominant red horizontal branch morphology like most of the Galaxy's
dSph companions.  The results were interpreted as indicating that the 
bulk of And~I's stellar population was $\sim$10 Gyr old.  The presence of blue
horizontal branch and RR~Lyrae stars, however, testified to the presence of
an older population in And~I\@.  Thus, like most of the Galactic dSph
companions, And~I has had an extended epoch of star formation.
The WFPC2 data also revealed a radial gradient in And~I's
horizontal branch (HB) morphology, in the sense that there are relatively more
blue HB stars beyond the dSph's core radius.  This was interpreted  
as evidence that the subsequent star formation was more centrally
concentrated than the initial episode.
In this paper we concentrate on And~II\@.  Observations
of And~III with WFPC2 were obtained recently and they will be the subject
of the next paper in this series.  The newly discovered
systems And~V and And~VI are scheduled to be observed in a Cycle 8 HST General
Observer program. 

And~II is one of the more luminous M31 dSph companions and, aside from And~III, it has the largest apparent flattening of these systems 
(\markcite{NC92}Caldwell \etal 1992, \markcite{NC99}Caldwell 1999).  
It lies $\sim$140 kpc in projection from the center of M31
in the direction of M33.  Indeed, pending the results presented in this paper,
it is possible that And~II is associated with M33 rather than M31.  And~II
is unique among the M31 dSph satellites in that it is the only one in
which the existence of upper-AGB carbon stars has been spectroscopically 
confirmed (\markcite{AM85}Aaronson \etal 1985, see also 
\markcite{CO99}C\^{o}t\'{e} \etal 1999b).  
These stars are relatively common in the Galactic dSphs and are an indicator
of the presence of an intermediate-age ($\sim$1 -- 10 Gyr) population.
Previous ground-based studies of And~II include that of 
\markcite{KN93}K\"{o}nig \etal (1993)
in which it was suggested that And~II was somewhat closer along the
line-of-sight than M31, and
that this dSph perhaps has a relatively large intrinsic metallicity spread.
Our WFPC2 data allow a more precise and quantitative evaluation of these
possibilities.  C\^{o}t\'{e} \etal (1999b) have also investigated the mean
metallicity and intrinsic metallicity spread using (ground-based) low
resolution spectra of a sample of And~II red giants.

The remainder of this paper is split into three sections.  In 
Sect.\ \ref{O_Rsect} we discuss the observations, the photometric techniques
employed and the calibration process adopted.  In Sect.\ \ref{results}, the
WFPC2 c-m diagrams are presented and main results of the paper
discussed.  These include the HB morphology, the true distance of And~II
from the center of M31, the mean abundance and abundance dispersion, and
inferences on the age(s) of the stellar population.  The 
consequences of our results in the wider context of dSph evolution
are discussed in the final section.  An Appendix describes minor revisions
to the And~I results of Paper~I made necessary by the reduction and analysis 
techniques adopted in this paper.

\section{Observations and Reductions} \label{O_Rsect}

Andromeda II was imaged with the WFPC2 instrument aboard the {\it Hubble Space
Telescope} on 1997 August 29 and again, at the same orientation, on 1997
September 3 as per our GO Program 6514.  The first set of observations 
consisted of three 1200~s
integrations through the $F555W$ (``Wide-$V$'') filter and seven 1300~s
integrations through the $F450W$ (``Wide-$B$'') filter.  The second set of
observations used identical exposure times but there were four $F555W$
and eight $F450W$ integrations.  The field of And~II contains a number of
bright foreground stars (e.g.\ Caldwell \etal 1992, Fig.\ 2).  
Consequently, to avoid any possible deleterious effects from these stars, 
a WFALL-FIX center offset from the center of the dSph, and a range of 
allowable orientations, was specified at the Phase-II stage.  The actual 
observations placed
the center of And~II approximately on the boundary between the WF3 and WF4 
chips $\sim$70$\arcsec$ from the WFPC2 pyramid apex.  The second set of
observations was displaced from this center by a small amount, nominally 
9.5 WF pixels in both $x$ and $y$.  This facilitates distinguishing real stars
from instrumental defects, such as hot pixels, and allows an assessment of 
the errors in the photometry.  In this respect these And~II observations
are similar to those of And~I described in Paper~I.  The location of the
section of And~II imaged with WFPC2, relative to the full extent of this dSph 
galaxy, is shown in Fig.\ \ref{wiynpic}.

The raw frames were processed via standard STScI pipeline procedures.  In
particular, the DARKCORR (dark correction) frame used, which includes the 
(time variable) hot pixels, was obtained 
just 4 days before the first set of observations.  The 
processed frames were then separated into the images for each individual CCD,
multiplied by the appropriate geometric distortion image
as supplied by STScI, and trimmed of the vignetted regions using the 
boundaries recommended in the
WFPC2 Handbook.  The locations of bright stars were then measured
on each set (one for each combination of filter, position and CCD) to
ascertain whether there were any systematic changes in position
during the sequence of observations.  None were found.  The individual
images for each set were then combined using the $gcombine$ task
within the STSDAS package.  This task removes cosmic rays through 
a sigma-rejection
criterion but one must be careful that the centers of bright stars are not
adversely affected at the same time.  We experimented with the parameter
$snoise$ until we found values (typically $\sim$0.10) that gave excellent
cosmic ray rejection but for which magnitudes of bright stars on the
combined frame were not significantly different ($\lesssim$0.01 mag)
from the average of the values
from the individual frames.  A mosaic made from the combination of four 1200~s 
$F555W$ frames is shown in Fig.\ \ref{wfpc2pic}.  As for the And~I data in
Paper~I, the HST/WFPC2 combination completely resolves this dSph; indeed
the And~II stars are relatively uncrowded.  There is also no indication of any
star clusters on this image (nor are there any candidates from 
ground-based imaging), a result that is not surprising given that And~II
is $\sim$2 magnitudes fainter than the least luminous dSph galaxy with
an identified star cluster population.

\subsection{Photometry}

The relatively uncrowded nature of the And~II frames means that we can
employ the techniques of aperture photometry.  We note first, however, that 
because 
of the small number of stars on the PC1 frames ($\sim$5\% of the total number 
on the WF chips), and because of the lack of ``bright'' stars with
which to determine the PC1
aperture corrections, we have decided not to use the PC1 data.  
All subsequent discussion then applies solely to the three WF CCDs.
Briefly, the {\it daofind}
routine within IRAF/DAOPHOT was used to generate an image-center coordinate
list for each of the 12 combinations of filter, position and WF frame.  
Aperture photometry using a 2 pixel radius aperture was then carried
out for the objects on the input lists, with the ``sky'' taken as the mode 
of the pixel values in an annulus of inner radius 5 and outer radius 15 pixels.
This was then followed by the determination of the aperture corrections -- the
difference between the 2 pixel measurement aperture and the 5 pixel
standard aperture adopted by \markcite{H95B}Holtzman \etal (1995b, 
hereafter H95).  Typically a dozen or
so of the brightest, most isolated stars spread across each WF image were used.  The aperture corrections for
the individual stars, which showed a standard 
deviation of typically 0.02 mag, were then averaged to form a single correction 
for each filter/position/frame combination.  The corrected photometry lists 
for the two positions and a given filter
were then compared.  There were no indications of any systematic differences
and as a result, a single magnitude was generated by averaging the two measurements.  Stars detected 
on only one frame were discarded.  Similarly,
the combined photometry lists for the $F555W$ and $F450W$ filters were 
matched to produce $F450W-F555W$ colors.  Again stars not detected in both
filters were ignored.  
While this does limit the data to those stars detected on the ``shallowest''
frames (the 3 $\times$ 1200 s $F555W$ combination for bluer stars and
the 7 $\times$ 1300 s $F450W$ combination for redder stars) it does ensure
that the objects in the photometry lists are real.  At this stage corrections 
for exposure time, gain factors and zeropoint were also applied to  
place the photometry on the H95 system.

Although the frames are relatively uncrowded, there are still occasions
where the signal in the measurement aperture for a star is contaminated by
signal from a nearby companion, leading to an (undetected) error in the
measured magnitude and to an increase in the scatter in the c-m diagram.  
To avoid this possibility, we have removed from the photometry
lists any star whose center lay within 5 pixels of the center of another star.
This reduced the sample by less than 10\% but reduced the scatter
in the c-m diagram.
Finally, the remaining ``stars'' were visually inspected on the 
four-exposure-combined $F555W$ images and a small number of objects, 
mostly marginally resolved galaxies, removed.  In these respects the procedures
followed here are identical to those used for And~I in Paper~I\@.  
The final And~II sample then consists of 1765 stars from the three WF frames.

\subsubsection{Charge-Transfer Efficiency Corrections} \label{cte_sect}

The WFPC2 CCDs are known (e.g.\ \markcite{H95A}Holtzman \etal 1995a) to suffer
from poor charge-transfer efficiency (CTE) which can affect photometry
obtained from WFPC2 images, especially for faint stars on exposures with a
low background (sky) level.  \markcite{ST98}Stetson (1998) gives an excellent
description of the likely physical processes involved.  Holtzman \etal (1995a)
proposed for low background situations, as in our exposures where the
sky level is typically $\sim$65 $e^{-}$pixel$^{-1}$ for $F555W$ and
$\sim$30 $e^{-}$pixel$^{-1}$ for $F450W$, that 
the effect of the poor CTE could be 
reduced by imposing a 0.04 mag/800 pixel ramp correction, 
in the sense that stars with higher $y$ coordinate become
relatively brighter.  This correction process was adopted in Paper I\@.
More recently, however, \markcite{WH97}Whitmore \& Heyer (1997, 
hereafter WH97), and subsequently Stetson (1998), have investigated 
the CTE effects in greater detail.
Both report that the CTE effects are a function of not only the $y$, but also
the $x$, coordinate on the frames.  WH97 give a variety of
formulae to correct for the effects.  We applied their equations 2c and
3c (2 pixel radius aperture, local sky) to first correct the individual 2 pixel
aperture magnitudes on all the filter/position/WF frame photometry lists.  
We then used WH97 equations 2d and 3d (5 pixel radius aperture, local sky) to correct the magnitudes of the aperture correction stars and thus
generate new aperture corrections.  We then regenerated 
the final averaged, crowding corrected, edited and calibrated photometry list.
For our data, the WH97 CTE correction process
makes a typical And~II red horizontal branch star near the center of a WF
frame brighter by $\sim$0.04 mag and bluer by $\sim$0.01 mag relative to the
uncorrected photometry; the effects are about twice as large for
stars that have large $x$ and $y$ coordinates.  Stetson (1998) indicates that his formalism generally results in similar
corrections to those of WH97, so we have chosen not to investigate the
application of Stetson's formalism to our data.

\subsubsection{Photometric Errors}

With the CTE corrected data set we can now investigate the photometric errors,
i.e.\ the errors that arise from the measurement process, as distinct from
systematic errors that can come from uncertainty in the aperture corrections,
in the zeropoint calibration, in the CTE correction process and so on.  To do
this we return to the photometry lists for the two separate pointings and 
compare colors and magnitudes for those stars in the final sample.
The results of this process are presented in Table \ref{error_tab}.  It 
lists, for the specified magnitude bins, the average error in the mean of the
two measures (either magnitude or color).  As was found in Paper~I, at fainter 
magnitudes for these uncrowded stars, the photometric errors are essentially 
those expected on the basis of photon statistics errors alone.  The 
errors themselves though are somewhat larger than in Paper~I 
as a consequence of the
shorter total integration times.  At brighter magnitudes, however, the
photometric errors approach a lower limit of $\sim$0.015 mag, again
as was found in Paper~I\@.  We ascribe this limit on the photometric precision
to the effects of flat fielding and dark subtraction, and to the individual
frame combination process adopted.  The existence of this limit, however, 
does not affect the interpretation of the data in any way.  

In the comparison process a small number of stars were identified as 
having $F555W$, $F450W$ or $F450W-F555W$ differences larger by 3.5$\sigma$ 
or more, than the mean difference for their magnitude (which was always
close to zero).  As we shall see 
in Sec.\ \ref{rrvar}, the majority of these stars are likely to be 
And~II RR~Lyrae variable stars.  Fig.\ \ref{wfpc2cmd} then shows 
the color-magnitude
diagram for the final And~II sample on the HST/WFPC2 photometric system of H95.  The 54 candidate variable stars are plotted with a different symbol.   

\subsubsection{$V$, $B-V$ Zeropoint Calibration} \label{bv_zpt}

\begin{figure}[p]
\plotone{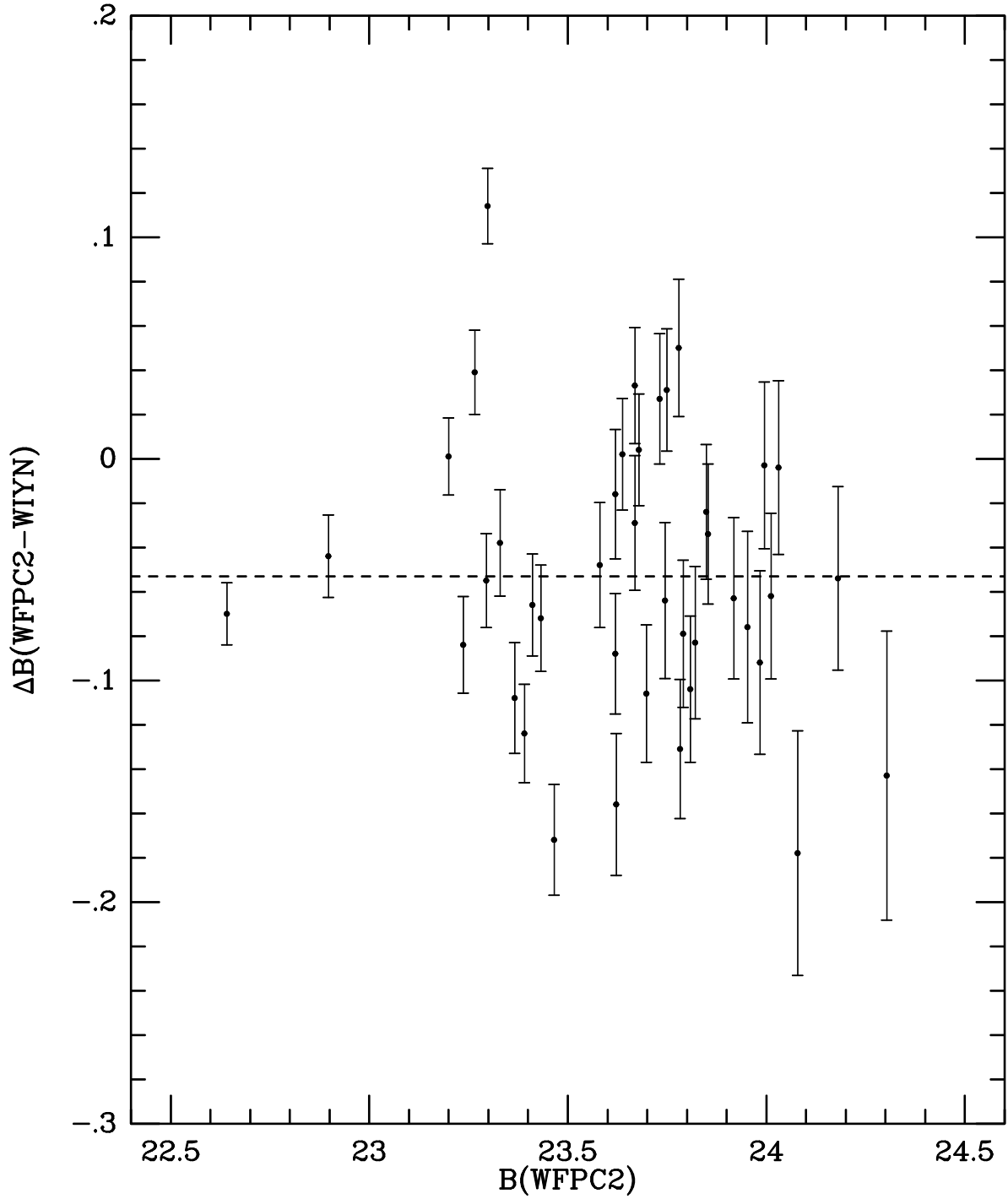}
\caption[fig4.ps]{A plot of the difference between $B_{WFPC2}$,
derived from the $F450W$ photometry via application of the
transformations in Holtzman \etal (1995b), and $B_{WIYN}$, derived
from ground-based imaging, against $B_{WFPC2}$ for And~II red giants.  The
mean offset is --0.053 $\pm$ 0.010 mag.  \label{wiynb} }
\end{figure}

In paper I we adopted the $F555W$ to $V$ and $F450W$ to $B$ transformations as
given in H95.  For $F555W$ the transformation is derived from ground-based
observations using a replica of the flight system (see H95 for details) but
for $F450W$ the transformation is not directly determined.  Instead it comes
from the convolution of the HST+WFPC2+CCD+$F450W$ filter response curve with
a library of stellar spectra.  It is then (perhaps) less well determined
than that for $F555W$.  In order to investigate these transformations,
or more particularly their zeropoints, we obtained 
a set of deep $B$ and $V$ images of And~II with the 
WIYN\footnote{The WIYN Observatory is a 
joint facility of the University of Wisconsin-Madison, Indiana University, 
Yale University and the National Optical Astronomy Observatories.}
3.5 m telescope via the NOAO WIYN Queue process.  These images
were taken in excellent seeing (FWHM $\approx$ 0.55$\arcsec$) and under
photometric conditions.
A number of photometric standard star fields 
were also observed to provide calibration.  These photometric standard
observations generated very well determined 
transformations from the WIYN instrumental system to the $B, V$ standard system.
We then inspected 
the WFPC2 and WIYN images and selected $\sim$40 stars distributed over the WF 
frames that could be reliably
measured on the WIYN images.  By necessity, these stars are all at or near
the top of the And~II red giant branch.  The differences between the WFPC2
and the WIYN photometry for these stars were then derived.  Considering first 
the comparison
of $V_{WFPC2}$, which arises from the CTE corrected $F555W$ photometry, and 
$V_{WIYN}$, we find no indication of any offset, nor any trend with magnitude
or color.  For the 41 stars involved in this comparison, the
mean value of $V_{WFPC2}-V_{WIYN}$ is 0.007 mag with a standard error in
this mean of 0.010 mag.  We conclude, as have many others, that the 
$F555W$ to $V$ transformation is well determined.  The situation for 
$F450W$ and $B$, however, is not as good.  As illustrated in Fig.\ \ref{wiynb}
we find that there is a clear offset between $B_{WFPC2}$, which arises 
from the CTE corrected $F450W$ photometry and the H95 synthetic 
transformation, and 
$B_{WIYN}$.  The mean value of $B_{WFPC2}-B_{WIYN}$ is --0.053 mag with a 
standard error in this mean of 0.010 mag.  There does not seem to be any
indication that this offset varies with color, though the range available
is limited (1.0 $\lesssim$ $B-V$ $\lesssim$ 1.7).  

To confirm this
apparent zeropoint offset, we searched the HST archive for additional
observations which employed the $F450W$ filter and which targeted objects
with established $B$, $V$ photometry.  The WFPC2 Cycle 6 Calibration Program
6935 proved ideal for this test, since it contains $F450W$ and $F555W$
observations of the $\omega$ Cen standard star field.  Ground-based
$B$, $V$ photometry is available for this field from 
\markcite{AW94}Walker (1994).  We performed photometry on the WFPC2 frames 
of the $\omega$ Cen standard star field 
in exactly the same way as was done for And~II, including the application
of the WH97 CTE corrections.  This led to 42 measures of 20 standards
obtained from three different WFPC2 data sets (different orientations meant
that not all stars are found on all frames).  The magnitudes and colors for 
the multiply observed stars were averaged and the WFPC2 data for all
compared with the Walker (1994) ground-based values.  Once again 
we find no strong evidence for any
systematic differences when comparing $V_{WFPC2}$ with $V_{Walker}$.  For
the $B-V$ colors, however, an offset almost identical in size to that 
found for And~II is present.  Weighting by the number of observations,
the mean value of $(B-V)_{WFPC2}-(B-V)_{Walker}$ is --0.049 mag
with a standard error in the mean of 0.006 mag, for 38 observations of 18
stars (we have excluded the three observations of the sole blue star and one
notably discrepant measure).
The majority of the stars in this comparison are $\omega$ Cen main sequence
stars and are thus bluer in color than the And~II red giants.  Yet, as for
And~II, there is no evidence for any correlation between the difference
$(B-V)_{WFPC2}-(B-V)_{Walker}$ and $(B-V)_{Walker}$, at least for $(B-V)$
colors redder than $(B-V)$ $\approx$ 0.50 (the reddest star in the $\omega$
Cen sample has $B-V$ = 1.33 and thus there is color overlap with the 
And~II sample).  As to whether this offset 
applies for bluer colors, we lack sufficient information to be certain one
way or the other.  We assume that it does but, fortunately, no results in this paper are affected by this latter assumption. 

Thus, in transforming the And~II $F450W$, $F555W$ CTE corrected photometry 
to the standard $B$,$V$ system, we have adopted
a zeropoint shift of 0.055 mag relative to the H95 relation, in the sense 
of fainter $B$ magnitudes and thus redder $B-V$ colors.  The resulting c-m diagram, without the candidate variables, is shown in Fig.\ \ref{bv_cmd}.
The implications of this change in adopted $B-V$ zeropoint, and of 
the WH97 CTE corrections, for the And~I results presented in Paper~I are
discussed in Appendix A.

\section{Results}   \label{results}

\subsection{The Color-Magnitude Diagrams of And II}

In general, the c-m diagrams for And~II presented in Figs.\ \ref{wfpc2cmd} 
and \ref{bv_cmd} bear a strong resemblance to those for And~I in
Paper~I (and in Appendix~A).  The basic morphology is that of an old
stellar population -- there is a red giant branch with a notable color width,
especially at brighter magnitudes, and a horizontal branch showing a large
color range but with a predominance of red horizontal branch stars.  A 
population of stars fainter than the horizontal branch and bluer than the
giant branch is also evident.  The most conspicuous difference between the
And~I and And~II c-m diagrams, however, is the relative lack of objects redder 
than the ``red edge of the giant branch'' in the And~II c-m diagram.  In 
Paper~I we argued that such stars in the And~I c-m diagram were predominantly
red giants from the halo of M31; red Galactic foreground stars
were approximately a factor of two less important.  That assertion in Paper~I
is supported here -- since And~II lies $\sim$3 times further in projection from 
the center of M31 than does And~I, we expect the And~II c-m diagram to be
much less contaminated by M31 halo stars, and this is what we see.  
Indeed, using the predictions of \markcite{RB85}Ratnatunga \& Bahcall (1985) 
as a guide, we can suggest that the small
number of redder objects in Figs.\ \ref{wfpc2cmd} and \ref{bv_cmd} are
probably all Galactic foreground objects\footnote{One definite exception to
this statement is the star that lies $\sim$0.4 mag fainter but at the same
color as the brightest, reddest giants in Figs.\ \ref{wfpc2cmd} and 
\ref{bv_cmd}.  This is C\^{o}t\'{e} \etal (1999b) star 3; a spectroscopically identified carbon star that is a highly probable And~II member.}.
Consequently, given that the color
distribution of Galactic foreground objects greatly favors red stars at these
magnitudes (e.g.\ Ratnatunga \& Bahcall 1985), we conclude
that the degree of contamination of the And~II c-m diagrams by non-member
objects is negligible.

\subsubsection{The Horizontal Branch Morphology} \label{hbmorph}

It is clear from Figs.\ \ref{wfpc2cmd} and \ref{bv_cmd} that the morphology
of the horizontal branch in And~II is dominated by red stars.  As was
the case for And~I, the red side of the And~II HB is superposed on the
red giant branch and there is no clear distinction between the red giants and
the core helium burning stars at this luminosity.  
In Paper~I we noted that this is a
consequence of three effects: a spread in the colors of 
HB and red giant stars of similar luminosity due to photometric errors
in $F450W-F555W$; the lower sensitivity
(relative to $B-V$) of $F450W-F555W$ to differences in effective temperature;
and the presence (see Sect. \ref{abund_spd}) of a significant metal abundance 
spread, which results in horizontal branch stars from the more metal-rich
population having comparable colors to similar luminosity red giants from the
metal-poor population.  These effects are no doubt also responsible for the
appearance of Figs.\ \ref{wfpc2cmd} and \ref{bv_cmd} at HB luminosities.

To quantify the And~II HB morphology we follow Paper~I and calculate the HB
morphology index $i$ = $b/(b+r)$, where $b$ and $r$ are the number of blue
and red HB stars, respectively.  In particular, we take $r$ as the number
of stars with 24.65 $\le$ $F555W$ $\le$ 25.15 and 0.35 $\le$ $F450W-F555W$
$\le$ 0.60.  The red limit was fixed by noting that a
color histogram for stars in this magnitude range shows a decline redward
of this color, and thus it represents a reasonable estimate for the color
limit beyond
which red giants outnumber red HB stars.  The same limit was adopted in 
Paper~I\@.  For $b$, we count the stars with 24.65 $\le$ $F555W$ $\le$ 25.60
and --0.35 $\le$ $F450W-F555W$ $\le$ 0.25, less the small number of stars
with 25.15 $\le$ $F555W$ $\le$ 25.60 and 0.10 $\le$ $F450W-F555W$ $\le$ 0.25
since they are probably not blue HB stars.  Including the candidate variables,
we find $r$ = 343 and $b$ = 76 for $i$ = 0.18 $\pm$ 0.02, where the error is
calculated assuming that $b$ and $r$ follow Poisson statistics.  If we
exclude the candidate variables, then the value of $i$, 0.17 $\pm$ 0.02, is
essentially unchanged.  

For And~I we found $i$ = 0.13 $\pm$ 0.01 suggesting
that the And~II HB morphology is somewhat bluer than that of And~I\@.  In
quantitative terms, if And~II had possessed the same HB morphology index 
as And~I, then for the same total number of HB stars as observed we would have expected $\sim$50 blue HB stars.  In fact approximately 50\% more than
this are present.  This result does not depend on the adopted red limit for 
defining the red HB population.  If instead of $F450W-F555W$ = 0.60, we adopt 
0.52 for this limit, the absolute values of the morphology index $i$ change 
but the result that And~II contains $\sim$50\% more blue HB stars relative to
And~I, remains.   

Despite this difference though, the And~II HB morphology, like that of And~I, is
dominated by red HB stars.  Standard Galactic globular clusters whose
abundances encompass the And~II mean ($<$[Fe/H]$>$ $\approx$ --1.5, see Sect.\ 
\ref{abund}) have horizontal branch morphologies that
show an even, or a blueward biased, distribution of HB stars.  For example,
using the R and B numbers given in \markcite{YL89}Lee \etal (1994), we find 
$i$ = 0.74 $\pm$ 0.04
for M5 ([Fe/H] = --1.40) and $i$ = 0.45 $\pm$ 0.05 for M4 ([Fe/H] = --1.28)
while the clusters NGC~6752 ([Fe/H] = --1.54) and M13 ([Fe/H] = --1.65)
have $i$ = 1.00, i.e.\ entirely blue HB morphologies.
At first sight this difference between the value of $i$ = 0.18 $\pm$ 0.02
for And~II and those for these standard Galactic globular clusters, 
suggests that And~II should be classified
as showing the {\it second parameter effect} in the same way as does And~I 
(Paper~I).  For instance, the well known ``second parameter'' Galactic
globular cluster NGC~362 ([Fe/H] = --1.28) has a dominant red HB morphology characterized by $i$ = 0.04 $\pm$ 0.02, while NGC~7006 ([Fe/H] = --1.59), also
a second parameter cluster, has $i$ = 0.23 $\pm$ 0.06.  
However, the large abundance range in And~II (see Sect.\ \ref{abund_spd})
means that any conclusion regarding the second parameter nature of And~II
that is based solely on these simple comparisons must be viewed with 
some caution.  
Nevertheless, we shall show in Sect.\ \ref{age_sect} that the 
HB morphology of And~II
cannot be modelled with a combination of standard (i.e.\ non second
parameter) Galactic globular
cluster horizontal branches -- a significant second parameter component 
to the HB morphology is required.  Thus we
are entitled to claim that And~II is a second parameter object along with
And~I and most of the Galactic dSph companions.

One of the most intriguing results of Paper~I was the discovery 
that And~I shows a radial
gradient in its HB morphology, in the sense that there are relatively more
blue HB stars outside the dSph's core radius.  Similar HB morphology gradients
also exist in at least two Galactic dSphs but are absent in others (Paper~I).
Consequently, we have used our data to investigate whether or not such a
HB morphology gradient is present in And~II\@.  According to 
Caldwell \etal (1992), And~II has an ellipticity of 0.3 and
a (geometric mean) core radius of $\sim$100$\arcsec$ ($\sim$120$\arcsec$ on the
major axis).  We first split our And~II photometry sample into inside and
outside the core radius samples, using the appropriate elliptical boundary
(the position angle of the major axis for And~II is $\sim$155$\arcdeg$). 
The average radial distances of the stars in these samples are 58$\arcsec$
and 130$\arcsec$, respectively, with the most distant stars $\sim$170$\arcsec$
from the center.  We
also split the sample, again along an elliptical boundary, into two groups
with approximately equal numbers of stars.  This occurs for a major axis
radius of $\sim$75$\arcsec$ or $\sim$0.6r$_{core}$.  The average radial
distances for these samples are 42$\arcsec$ and 99$\arcsec$, respectively.

We find, using the 
sample which includes the candidate variable stars, that inside the core
radius $i$ = 0.18 $\pm$ 0.02 ($r$=289, $b$=62) while outside the core
$i$ = 0.21 $\pm$ 0.05 (54, 14).  Similarly, inside the $\sim$50\% sample
radius, $i$ = 0.16 $\pm$ 0.03 (180, 35) while outside this radius
$i$ = 0.20 $\pm$ 0.03 (163, 41).  None of these differences are statistically
significant (recall from Paper~I that the value of $i$ for And~I changed from 
$i$ = 0.11 $\pm$ 0.01 inside the core radius to $i$ = 0.20 $\pm$ 0.03 outside
it, a 2.9$\sigma$ difference).  We conclude then that And~II apparently
differs from And~I in that it lacks any significant HB morphology radial 
gradient, at least within $\sim$1.3 core radii.  
The existence of this difference should not be seen as too 
surprising since, as we noted in Paper~I, among the Galactic dSphs Leo~II
and Sculptor possess similar HB morphology gradients to And~I but Carina 
does not.

\subsubsection{RR Lyrae Variables} \label{rrvar}

\begin{figure}[p]
\plotfiddle{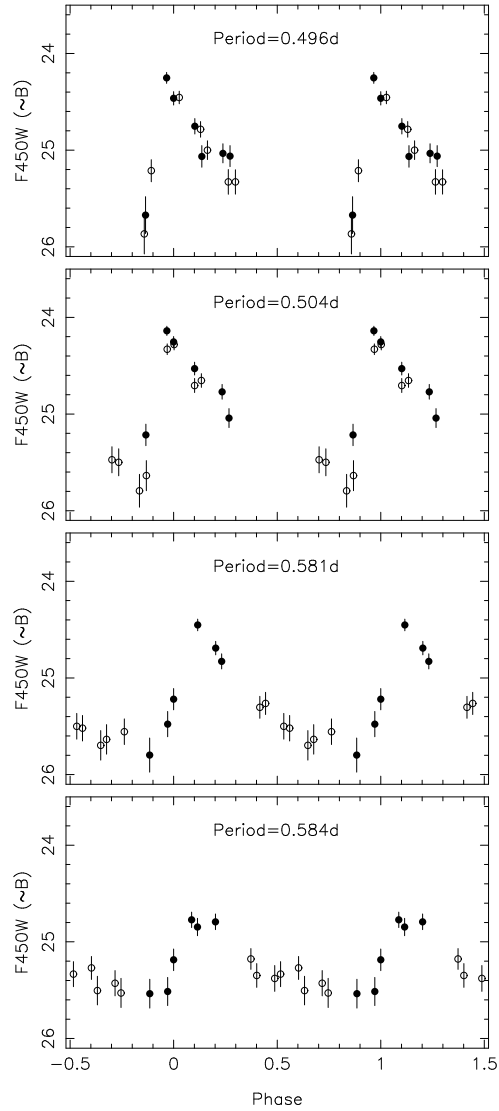}{7.3in}{0.}{60.}{60.}{-210.}{-15.}
\caption[fig6.ps]{$F450W$ light curves for 4 And~II RR~Lyrae stars.
These stars come from the sample of candidate variables identified in 
Fig.\ 3.  Two cycles are plotted for each variable with open
and closed symbols representing the two sets of observations.  The error bars
are those from the photon statistics.  The adopted periods are given in
each panel.  \label{vars_lc} }
\end{figure}

In Fig.\ \ref{wfpc2cmd} there are approximately 30 candidate variable stars
with magnitudes similar to that of the horizontal branch.  These stars are
most likely to be RR~Lyrae variables in And~II\@.  To investigate  
this assertion we have performed aperture photometry for a subset of these
candidates in the same way as discussed above, except using the individual 
WFPC2 frames rather than the combinations.  This process yields, cosmic-ray 
contamination willing, a maximum of 7 individual $F555W$ and 15 $F450W$ 
magnitudes for each candidate.  A plot of these magnitudes against the 
mid-exposure Julian date then clearly indicates whether or not the 
candidate varies in a systematic way.  Of the 13 candidates investigated
in this fashion, all showed clear evidence of variability with timescales and
amplitudes typical for RR~Lyrae stars.  Attempts were then made to find
periods for these stars.  However, the relatively short duration of the two 
sets of observations, $\sim$0.2 d, compared to the interval between them, 
$\sim$4.7 d, often meant that it was difficult to choose between two
distinct periods which gave similar light curves (typically resulting from
the ambiguity as to whether there were 8 or 9 cycles between the two sets
of observations).  The $F555W$ observations were frequently used (along with
the assumption of $\sim$constant color) to reduce this ambiguity as was the
assumption that these RR~Lyraes should follow an approximate
period-amplitude relation.
Nevertheless, it was often not possible to determine a unique period.
Typical light curves resulting from this process are shown in 
Fig.\ \ref{vars_lc}.  These light curves, and others not shown, identify
most of these stars as Type-ab RR Lyraes.  For one star it was not possible
to generate a reasonable light curve while another may be a Type-c variable
with a lower amplitude and shorter period.

In this process it was noticed that two of the five HB stars with extremely
blue $F450W-F555W$ colors in Fig. \ref{wfpc2cmd} are RR~Lyrae variables,
while the other three such stars were not flagged as variable candidates.  
Consequently, these three stars were also investigated for variability in the
same way as the other candidates.  Perhaps not surprisingly, 
they were also found
to be RR~Lyrae variables.  This demonstrates that the method of identifying
variables from significant differences in the mean magnitudes or colors 
between the two data sets, while successful, is not likely to generate a
complete set of variables.  A full investigation of the frequency of RR~Lyrae
variables on the And~II horizontal branch, and a discussion of the nature of
the non-HB
candidate variables identified in Fig.\ \ref{wfpc2cmd}, will be presented 
in a subsequent paper.

\subsubsection{The Giant Branch Intrinsic Color Width} \label{clr_wid}

One of the most notable results in the ground-based study of And~II by 
K\"{o}nig \etal (1993) was their suggestion that And~II has a 
large intrinsic color spread on the giant branch, and thus a significant
internal abundance range.  In particular, K\"{o}nig \etal (1993) measured
$\sigma_{obs}(g-r)$ = 0.20 on the upper giant branch in their c-m diagram.
They did not perform a quantitative analysis of their photometric errors but
instead estimated them as $\sigma_{err}(g-r)$ $\approx$ 0.15 mag.  This in turn
yields $\sigma_{int}(g-r)$ $\approx$ 0.13 for the intrinsic color dispersion,
which they estimate corresponds to $\sigma_{int}({\rm[Fe/H]})$ $\approx$ 
0.43 dex.  This value is amongst the largest known for dSph galaxies 
(e.g.\ \markcite{MM98}Mateo 1998).
Clearly our much smaller photometric errors 
($\sigma_{err}(F450W-F555W)$ $\approx$ 0.02 at comparable magnitudes) allow us
to place much tighter constraints on any intrinsic color range among the
giants in this dSph galaxy.  That is the subject of this section; we defer
to Sect.\ \ref{abund_spd} the discussion of the abundance range that 
results from our determination of the And~II giant branch intrinsic color spread.

\begin{figure}[p]
\plotone{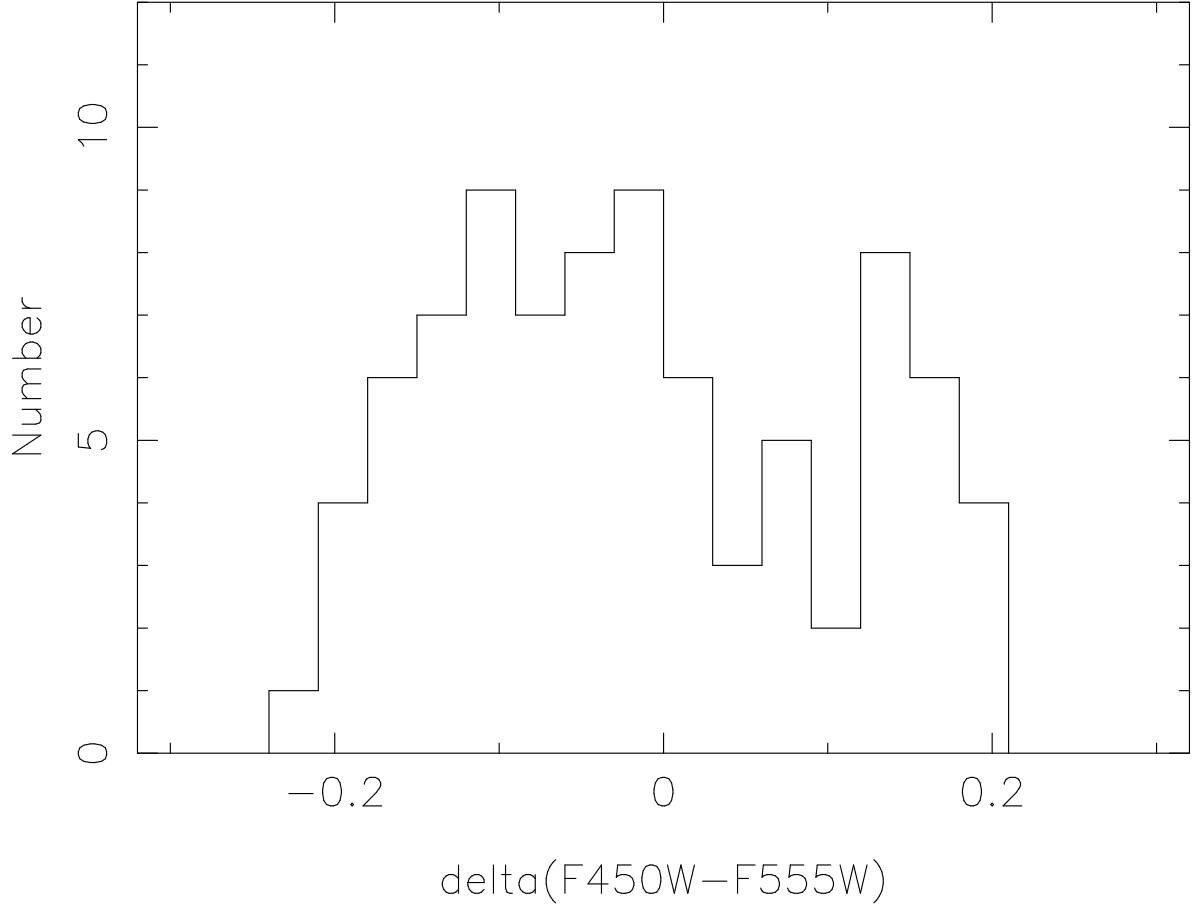}
\caption[fig7.ps]{Histogram of $F450W-F555W$ residuals from the
mean And~II giant branch for 85 stars with 22.2 $\leq$ $F555W$ $\leq$ 23.2.
The photometric errors for stars in this magnitude range is 
$\sigma_{err}(F450W-F555W)$ $\approx$ 0.02 mag.  The bin size is 0.03 mag. 
\label{clr_sprd_fig} }
\end{figure}

K\"{o}nig \etal (1993) considered the giant branch color spread in And~II
over the interval 22.75 $\leq$ $g$ $\leq$ 23.5 which corresponds approximately
to the interval 22.2 $\leq$ $F555W$ $\leq$ 23.2.  To investigate the intrinsic
color spread in this range we first determined a mean giant branch locus by
fitting a low order polynomial to the stars in the interval
21.8 $\leq$ $F555W$ $\leq$ 23.8, excluding a small number of outliers.  Then
for each of the 85 stars in the interval 22.2 $\leq$ $F555W$ $\leq$ 23.2,
we computed the difference between the $F450W-F555W$ color of the star and
that of the mean giant branch at the star's $F555W$ magnitude.  The
distribution of these differences is shown in Fig.\ \ref{clr_sprd_fig}.  This
distribution can be characterized in a number of ways: the standard deviation
$\sigma_{obs}(F450W-F555W)$ is 0.12 mag, the inter-quartile range is 0.18 mag,
the color range containing the central two-thirds of the sample is 0.27 mag
and the full color range of the sample is $\sim$0.4 mag.  All these quantities
are considerably larger than the errors in the colors at these magnitudes 
($\sigma_{err}(F450W-F555W)$ $\approx$ 0.02, cf.\ Table \ref{error_tab}).
Indeed the errors are so much smaller than the observed color range at
these magnitudes that we can effectively ignore them and regard the
observed quantities as intrinsic.  Thus we confirm the results of 
K\"{o}nig \etal (1993) that the upper giant branch of And~II has a
significant intrinsic color width.  Further, the values of the quantities
which characterize the distribution shown in Fig.\ \ref{clr_sprd_fig} are
all approximately twice as large as the equivalent quantities for the 
And~I intrinsic giant branch color width determined, using similar techniques,
in Paper~I (and Appendix~A).   Thus, while And~I and And~II have similar 
total luminosities
(Caldwell \etal 1992) and similar mean abundances, this does not
apply to their intrinsic giant branch color spreads; that of And~II
is significantly larger.

There are two further comments that are worth making here.  First, as noted in
Paper~I, at these luminosities in Galactic globular clusters, the AGB has 
either terminated or is no longer distinguishable from the red giant branch.
Thus the large intrinsic color width of the And~II giant branch is not due
to a mixture of AGB and red giant branch stars.  Second, there is a hint
of bimodality in Fig.\ \ref{clr_sprd_fig}, which is also evident to some
extent in Figs.\ \ref{wfpc2cmd} and \ref{bv_cmd}, though statistically the
distribution in Fig.\ \ref{clr_sprd_fig} is not significantly different 
from a gaussian.  We shall return to this point in Sect.\ \ref{abund_spd}.  

\subsubsection{A Population of ``Faint Blue Stars''?}  \label{fbs_sect}

In Paper~I we drew attention to the existence in the And~I c-m diagram
(see also Appendix~A) of a population of faint ($V$ $\gtrsim$ 26.0) 
blue ($B-V$ $\lesssim$ 0.5) stars.  These objects are sufficiently blue that
they cannot be easily dismissed as the result of large color errors in
subgiant star photometry, and hence they were considered 
a {\it bona fide} And~I 
population.  There is, however, no compelling evidence for {\it any}
intermediate-age population in And~I (\markcite{MK90}Mould \& Kristian 1990,
Armandroff \etal 1993, \markcite{TA94}Armandroff 1994).  Thus, in Paper~I we
interpreted these And~I stars as blue stragglers, analogous to those 
seen in many
Galactic globular clusters, rather than as main sequence stars with ages 
of order 1.5 -- 2.5 Gyr.

Because of the shorter exposure times, the limiting magnitudes in Figs.\
\ref{wfpc2cmd} and \ref{bv_cmd} are not as faint as the equivalent data
for And~I (cf.\ Appendix~A).  Nevertheless, it is apparent from these
c-m diagrams that And~II also has a recognizable population of faint
blue stars.  In particular, Table \ref{error_tab} shows that at $F555W$
$\approx$ 25.8, the mean repeatability error in $F450W-F555W$ is approximately 
0.11 mag.  Thus displacing a subgiant with a true color of $F450W-F555W$
$\sim$0.6 at this magnitude into the region of the faint blue stars
requires a color error of 3 or 4$\sigma$.  It then seems unlikely that all of
the objects with 25.6 $\lesssim$ $V$ $\lesssim$ 26.0 and $(B-V)$ $\lesssim$
0.5 could be explained in this way.

Visually, when comparing Fig.\ \ref{wfpc2cmd} or Fig.\ \ref{bv_cmd} with
the equivalent And~I c-m diagrams, it appears that And~II may contain 
relatively more of these faint blue stars.  We now seek
to quantify this impression.  Using the data of Fig.\ \ref{bv_cmd} there
are 39 And~II stars with 25.55 $\leq$ $V$ $\leq$ 26.0 and $(B-V)$ $\leq$ 0.5
mag.  These stars have an average $B$ of 26.1 mag.  Because at these faint
magnitudes the degree of completeness is limited by the $F450W$ data
(rather than by $F555W$), it is necessary to compare the number of these
faint blue stars not with subgiants in the same $V$ mag range, but with
a group of subgiants that have a similar mean $B$ magnitude.  The interval 
25.1 $\leq$ $V$ $\leq$ 25.4 and 0.70 $\leq$ $(B-V)$ $\leq$ 1.05 satisfies
this requirement: the 87 stars in this interval have $<B>$ $\approx$ 26.1 mag. 
To first order then the selection process should be the same for 
both groups of stars.  If $FBS$
denotes the number of faint blue stars and $SG$ denotes the number of
subgiants, then the proportion of faint blue stars of the total ($FBS+SG$)
for And~II is 0.31 $\pm$ 0.04 assuming Poissonian statistics.

As we shall see in subsequent sections (and in Appendix~A), the reddenings 
and mean metal abundances of And~I and And~II are similar.  The mean
magnitude of the horizontal branch, however, is $\sim$0.3 mag fainter in
And~I than it is in And~II\@.  We can thus calculate a similar proportion
for the faint blue stars in And~I by adopting the same color limits as 
for And~II but shifting the $V$ magnitude limits 0.3 mag fainter.  We find,
using the And~I data discussed in Appendix~A, that $FBS$ = 49 and $SG$ = 186
stars.  The proportion of faint blue stars is then 
0.21 $\pm$ 0.03, again
assuming Poissonian statistics.  This lower value for And~I 
confirms the subjective 
visual impression that And~II has relatively more faint blue stars -- 
if the And~I proportion applied in And~II then we would have expected
$\sim$23 rather than the observed 39 stars.  But is this difference 
statistically significant?  As discussed in 
Paper~I, the appropriate statistic for comparing two proportions is $T_{2}$ 
(see Paper~I for the definition), which has a normal distribution with zero 
mean and unit variance.  Using the numbers given above, we find $T_{2}$ = 2.13
which implies that there is a less than 2\% chance that the two proportions
are drawn from the same underlying population.  This result is not 
substantially altered by minor changes in the adopted magnitude and color
intervals used to define the $FBS$ and $SG$ samples.

As a check on this result we adopt a slightly different approach.  On the
upper part of the giant branch where completeness issues are not a concern
(we used the intervals 23.0 $\leq$ $V$ $\leq$ 23.5 and 
23.5 $\leq$ $V$ $\leq$ 24.0 for And~II, and the equivalent 0.3 $V$ mag fainter
intervals for And~I; two intervals allowing a consistency check), the number ratio of And~II to And~I stars in the c-m diagrams is 0.54 $\pm$ 0.02.  
On this basis and using the 49 observed faint blue stars in And~I,
we expect 26 such stars in And~II\@.  In fact we observe 39 such stars, or
$\sim$50\% more than expected.  We conclude then that And~II does contain 
significantly ($\gtrsim$50\%) more blue stars, with 
M$_{V}$ $\gtrsim$ +1.2, than does And~I\@.  The nature of these stars
will be discussed in Sect.\ \ref{bss_age}.

\subsection{The Distance of And II}  \label{dist_sect}

In Fig.\ \ref{bv_cmd}, the mean $V$ magnitude of the 45 blue horizontal
branch stars with 24.7 $\leq$ $V$ $\leq$ 25.1 and 0.05 $\leq$ $B-V$ $\leq$ 0.40
is 24.93 $\pm$ 0.015 (standard error of mean).  This value is essentially
unaltered for modest changes in the adopted color interval.  We thus adopt
$V$(HB) = 24.93 $\pm$ 0.03 for And~II, where the uncertainty now includes 
the adopted error in the aperture corrections ($\pm$0.02) and in the
$F555W$ to $V$ transformation zeropoint ($\pm$0.02).  In Paper~I we found
$V$(HB) = 25.25 $\pm$ 0.04 for And~I\@.  Thus it is immediately apparent,
given similar reddenings and mean abundances, that And~II is $\sim$0.3 mag
closer than And~I along the line-of-sight.

To convert the value of $V$(HB) for And~II into a distance, however,
we need to adopt
both a reddening and a mean abundance, since on our preferred distance 
scale, the absolute magnitude of the horizontal branch varies with abundance
($M_{V}$(HB) = 0.17[Fe/H] + 0.82, cf.\ Paper~I)\@.  For the reddening we use
the new COBE/DIRBE plus IRAS/ISSA dust maps of \markcite{SF98}Schlegel \etal
(1998), which allow for extinction fluctuations on smaller angular scales
than the earlier work of \markcite{BH82}Burstein \& Heiles (1982).  At the
galactic latitude and longitude of And~II, the Schlegel \etal (1998) maps
yield E($B-V$) = 0.060 mag, somewhat higher than the value E($B-V$) = 0.035
mag given by the Burstein \& Heiles (1982) data.  We adopt 
E($B-V$) = 0.06 $\pm$ 0.01 and A$_{V}$ = 0.19 $\pm$ 0.03 for And~II\@.  The
metallicity of And~II is the subject of the next section; here we simply adopt
the result, $<$[Fe/H]$>$ $\approx$ --1.5 $\pm$ 0.15 for And~II, yielding
$M_{V}$(HB) = 0.57 $\pm$ 0.03 on our adopted distance scale.  These data
then give (m--M)$_{0}$ = 24.17 $\pm$ 0.06 for And~II corresponding to a
line-of-sight distance of 680 $\pm$ 20 kpc.

To compare this distance with that of M31 requires an M31 distance that
is on our adopted distance scale.  As discussed in Paper~I, there are two such
determinations which give somewhat different results.  The field halo
RR~Lyrae and red giant data yield (m--M)$_{0}$ = 24.40 $\pm$ 0.13 
(Paper~I; see also \markcite{LM93}Lee \etal 1993) corresponding to a distance 
of 760 $\pm$ 45 kpc.  Alternatively, using our adopted distance scale, the mean
horizontal branch magnitudes of the eight M31 globular clusters studied
by \markcite{FP96}Fusi Pecci \etal (1996) suggest an M31 modulus of 
(m--M)$_{0}$ = 24.64 $\pm$ 0.05, or a distance of 850 $\pm$ 20 kpc.
These results then place And~II either 80 $\pm$ 50 or 170 $\pm$ 30 kpc
in front of M31 along the line-of-sight.  Thus we adopt a relative And~II -- M31
line-of-sight distance of 125 $\pm$ 60 kpc, and note that a more precise value
awaits a reduction in the uncertainty in the M31 distance (on our adopted
scale).  Our result, however, is consistent with the less precise
ground-based results
of K\"{o}nig \etal (1993) who found, via fitting globular cluster giant
branches to their c-m diagram, that And~II was 120$_{-100}^{+125}$ kpc closer 
than their adopted M31 distance.  

On the sky And~II lies closer to M33 than it does to M31 (see, for example,
Fig.\ 2 of \markcite{AD99}Armandroff \& Da~Costa 1999) suggesting the
possibility that And~II is associated with that galaxy rather than with 
M31.  However, since M33 lies 0.2 to 0.3 mag beyond M31 (e.g.\ Lee \etal 1993) while And~II lies on M31's near side, the association between And~II and M31 
seems unequivocal.

For our M31 moduli, the projected distance of And~II from M31 lies between 
135 and 150 kpc.  Combining these estimates with the line-of-sight relative
distance then indicates that the true distance of And~II from the center
of M31 lies between $\sim$160 and $\sim$230 kpc.  The lower value is 
reminiscent of 
the Galactocentric distance of the Fornax dSph (R$_{\rm G}$ $\approx$ 140 
kpc) while the upper value is comparable to the Galactocentric distances
of the Leo dSphs (R$_{\rm G}$ $\approx$ 250 and 205 kpc, respectively).
The Leo dSphs are the most distant of the Galaxy's system of dSph satellites.
And~II, however, is probably not the outermost member of the M31 dSph
system since the newly discovered M31 dSph companion And~VI lies at least
$\sim$270 kpc (the projected distance) from the center of M31 
(Armandroff \etal 1999).  

\subsection{The Abundance of And II} \label{abund}

The determination of an abundance estimate for And~II from these data
requires a comparison with the giant branches of standard globular clusters
of known abundance.  Since there are no giant branch data on the
($F555W$, $F450W-F555W$) system, we must use the And~II data transformed
to the ($V$, $B-V$) system.  As noted in Sect.\ \ref{bv_zpt}, our ground-based
photometry suggests that the transformation zeropoint for the $B$
magnitudes, and hence the ($B-V$) colors, should be altered by 0.055 mag
relative to the value given in H95.  It is this zeropoint corrected photometry
(cf.\ Fig.\ \ref{bv_cmd}) that we adopt here.  The standard globular cluster
giant branches are those for the Galactic globular clusters M68 ([Fe/H]=--2.09),
M55 ([Fe/H]=--1.82), NGC~6752 ([Fe/H]=--1.54), NGC~362 ([Fe/H]=--1.28) and
47~Tuc ([Fe/H]=--0.71) as described in Paper~I\@.  In particular, the
globular cluster abundances are taken from \markcite{TA89}Armandroff (1989)
and are on the \markcite{ZW84} Zinn \& West (1984) system.  For And~II
we adopt the Schlegel \etal (1998) reddening E($B-V$) = 0.06 and use A$_{V}$ = 
0.19 mag.

Since we employ a distance scale that depends on abundance, the determination
of the mean abundance, via the comparison with standard globular
cluster giant branches, and the determination of the distance modulus, are
necessarily interconnected.  In brief, we first assumed an estimate for the
mean abundance, which then allows the calculation of an And~II apparent 
distance modulus.  This permits the globular cluster giant branches
to be overlaid on the And~II photometry and a new estimate for
the mean abundance can be determined (see below).  The new mean abundance 
then yields a new modulus and the cycle is repeated; convergence is rapidly 
achieved.  We show the result of this process in Fig.\ \ref{bv_cmd_gb}, in
which the ($V$, $B-V$) photometry of Fig.\ \ref{bv_cmd} is presented with
the standard globular cluster giant branches superposed using the final adopted
And~II modulus.

To determine the mean abundance we computed mean ($B-V$) colors for the
stars, excluding obvious outliers, that lie in a series of 0.2 $V$ mag 
wide bins.  There were six
bins on the upper giant branch between $V$ = 22.15 and $V$ = 23.35 and two
bins on the lower giant branch between $V$ = 25.15 and $V$ = 25.55.  In this
latter case the magnitudes 
are fainter than that of the horizontal branch.  As discussed in
Paper~I, the lower giant branch offers the advantage of a larger sample 
and freedom
from contamination by stars evolving from the horizontal branch.  However,
at these fainter magnitudes the photometric errors are larger and the 
possibility of contamination from non-member objects is greater.  The reduced
sensitivity to abundance also means that small color uncertainties can result
in large abundance errors.  On the other hand, at higher luminosities, while
sample sizes are smaller, the photometric errors are also smaller 
(cf.\ Table \ref{error_tab}).  Further, the range in color corresponding
to a given abundance interval is larger and thus any small systematic 
uncertainties in the colors become less important.  The largest potential 
disadvantage of using the upper giant branch, however, is contamination 
from asymptotic giant branch (AGB) stars, which are generally hotter than
red giant branch (RGB) stars of similar luminosity and abundance.  Fortunately,
as noted in Paper~I, in the c-m diagrams of the standard globular clusters
(see Paper~I for the individual references) AGB stars cease to be readily
distinguishable from RGB stars brighter than M$_{V}$ $\approx$ --1.5 or
$V$ $\approx$ 22.9 in And~II.

The abundances were determined from the mean colors for each bin by applying
the abundance calibration for that bin.  These calibrations were determined by 
least squares fits (linear for all but the three most luminous bins where
quadratic relations were used) to the colors of the giant branches of the
standard clusters at the $V$ magnitudes of the bin centers, and the known
abundances of the clusters.  
For the two lower luminosity bins, which contain samples of 60
and 80 stars, respectively, the mean abundance determined is $<$[Fe/H]$>$ = 
--1.32 $\pm$ 0.06, where the error given reflects solely the uncertainty 
in the abundance resulting from the statistical uncertainties in the mean
colors.  For the six bins on the upper giant branch, where the sample sizes
range from 12 to 26 stars per bin, the mean abundance determined is 
$<$[Fe/H]$>$ = --1.49 $\pm$ 0.04 dex.  Again the error here is just that 
resulting from the statistical uncertainty in the mean colors.  For these 
upper giant branch bins there is no sign 
of any trend between the individual $<$[Fe/H]$>$ values and $V$ indicating, 
as expected, that the inclusion of AGB stars has not biased these abundance estimates.  

The abundance error due to the statistical uncertainty in the mean colors
is, of course, not the only uncertainty in the abundance estimates.  A full
accounting should include the effects of uncertainty in the $(B-V)$ zeropoint,
which we take as $\pm$0.03 mag, uncertainty in the distance modulus, which
we take as $\pm$0.06 mag (i.e.\ we exclude any systematic uncertainty in our
adopted distance scale), together with uncertainty in the abundance
calibration process itself.  This latter uncertainty we take as the rms
deviation about the least squares fits, which are of order 0.05 dex for both
the upper and lower giant branch.  For the upper giant branch mean abundance
determination, no single uncertainty significantly exceeds any other and the
quadratic sum of the contributions is 0.11 dex.  That is, the upper giant
branch estimate of the mean abundance of And~II is $<$[Fe/H]$>$ = 
--1.49 $\pm$ 0.11 dex.  For the lower giant branch the $(B-V)$ zeropoint
uncertainty dominates and the total error is 0.20 dex; i.e.\ the lower
giant branch estimate of the mean abundance is $<$[Fe/H]$>$ = --1.32 $\pm$ 0.20
dex.

The difference, 0.17 dex, in the sense of a lower abundance mean found
from the upper giant branch, is reminiscent of a similar difference found 
for And~I in Paper~I (see also Appendix A).  We prefer to adopt the mean
abundance determined from the upper giant branch for the principal reason
that our calibration of the $(B-V)$ zeropoint, on which the abundance 
determination ultimately rests, is predominantly based on upper giant branch
stars (cf.\ Fig.\ \ref{wiynb}).  Further, at fainter magnitudes the fiducial 
giant branches for the standard globular clusters are not as precisely defined  
as they are at luminosities above the horizontal branch.  Thus,
when combined with the large change in abundance produced by small color
differences at these magnitudes ($\Delta$([Fe/H])/$\Delta(B-V)$ $\approx$ 5.3),
this effect might be sufficient to generate the systematic offset.

We therefore adopt $<$[Fe/H]$>$ = --1.49 $\pm$ 0.11 dex as our best estimate
of the mean abundance of And~II\@.  This value is in good agreement with the
less precise value of $<$[Fe/H]$>$ = --1.59 $_{-0.12}^{+0.44}$ determined
by K\"{o}nig \etal (1993) from their ground-based ($g$, $g-r$) photometry.  
It is also in very good agreement with the recent spectroscopic And~II mean 
abundance given by C\^{o}t\'{e} \etal (1999b).  These authors
used line indices measured from low resolution Keck spectra of $\sim$40 And~II
red giants to determine $<$[Fe/H]$>$ = --1.47 $\pm$ 0.19 dex.  
Our value for the And~II mean abundance is also closely similar to the new mean
abundance for And~I, $<$[Fe/H]$>$ = --1.46 $\pm$ 0.12 dex, derived in
Appendix~A\@.  As noted above, these M31 dSph companions have closely similar
luminosities (Caldwell \etal 1992). 

One final point deserves mention.  To investigate whether or not there might
be a radial abundance gradient in And~II, we split the data as described in 
Sect.\ \ref{hbmorph} and computed mean abundances for the radially
selected samples from the 
giant branch colors.  For both the samples inside and outside the core 
radius, and inside and outside the elliptical boundary that divides our total
sample approximately in half, we see no evidence to suggest the presence of
any abundance gradient.  The 3-$\sigma$ upper limit on the size of any such
gradient is $\sim$0.3 dex within $\sim$1.3 core radii; larger samples of 
red giants in the outer regions 
of this dSph are required to generate a stricter limit.  This result is 
consistent with those for other dSphs (e.g.\ Paper~I) -- in no dSph system
is the existence of a radial abundance gradient firmly established.

\subsection{The Abundance Spread in And II} \label{abund_spd}

In section \ref{clr_wid} we demonstrated that the And~II giant branch has
a significant intrinsic color width.  The abundance calibration established 
in the previous section can now be used to convert this intrinsic
color width into an abundance distribution.  Specifically, we make use of the 
108 stars in the 6 upper giant branch $V$ mag bins employed to determine the 
And~II mean abundance.  By applying the appropriate abundance calibration to 
each individual star in each bin, we are able to generate 108 individual 
[Fe/H] values.
However, before collecting these values into an abundance distribution, it is
first necessary to estimate the uncertainty in these individual 
values.  

In this $V$ magnitude range, Table \ref{error_tab} shows that the mean error
in the $F450W-F555W$ colors is $\leq$ 0.02 mag.  Then noting that 
$\Delta(B-V)$/$\Delta(F450W-F555W)$ $\approx$ 1.25
and that $\Delta[Fe/H]$/$\Delta(B-V)$ $\approx$ 2.5
for these giant branch stars, the uncertainty in the
individual abundance determinations is approximately 
$\sigma_{err}$([Fe/H])
$\lesssim$ 0.08 dex.  We have therefore adopted a bin size of 0.15 dex in
generating histograms for the And~II abundance
distribution (as inferred from these giant branch stars).
The distribution is shown as
the solid lines in the panels of Fig.\ \ref{AndI_II}, where the upper and 
lower panels of the figure show the
result of offsetting the origin of the abundance bins by one half of a bin
width.

The observed abundance distribution shown in Fig.\ \ref{AndI_II} can be 
characterized
in a number of ways.  For example, the standard deviation 
$\sigma_{obs}$([Fe/H]) is 0.36 dex, the inter-quartile range is 0.66 dex, the
abundance range shown by the central two-thirds of the sample is 0.80 dex
and the full abundance range is approximately 1.4 dex.  The contribution
from the photometric errors, $\sigma_{err}$([Fe/H]) $\approx$ 0.08 dex, is
negligible by comparison and thus we can take all these quantities as
intrinsic.  In particular, we have $\sigma_{int}$([Fe/H]) = 0.36 dex.  This
value is somewhat smaller than the value, 
$\sigma_{int}$([Fe/H]) $\sim$ 0.43 dex,
derived by K\"{o}nig \etal (1993) from their ground-based photometry of the
And~II giant branch, but given the much larger photometric errors in that 
study, the agreement is quite satisfactory.  Our result is also again in 
excellent agreement with that of C\^{o}t\'{e} \etal (1999b), who derive
$\sigma_{int}$([Fe/H]) = 0.35 $\pm$ 0.10 dex from the dispersion of 
Mg $b$ line strength indices in their low resolution spectra of approximately
40 And~II red giants.  These authors also estimate 
$\sigma_{int}$([Fe/H]) $\approx$ 0.46 $\pm$ 0.17 from the dispersion in their
$(V-I)$ giant branch colors.

How do these And~II results compare with the intrinsic abundance dispersions
in the other M31 dSph companions?  For the recently discovered systems
And~V, And~VI (Peg dSph) and And~VII (Cas dSph), the limited information available lacks sufficient precision for useful comparisons. 
Armandroff \etal (1999) indicate that
$\sigma_{int}$([Fe/H]) $\sim$ 0.3 dex for And VI based on their ground-based
$V,I$ photometry, although they did not carry out a full analysis of their
photometric errors.  On the other hand,
\markcite{GG99}Grebel \& Guhathakurta (1999)
indicate only that ``the observed width of the RGB'' in the Peg
(And~VI) and Cas (And~VII) dSphs ``corresponds to a metallicity spread of
roughly 1 dex''.  This leaves And~I (Paper~I, see also Appendix~A) 
and And~III (\markcite{AD93}Armandroff \etal 1993) as the only
suitable comparison objects.  For And~I
the analysis in Appendix~A was carried out using similar WFPC2 data and 
the same technique.  Although this dSph has a similar total luminosity
and central surface brightness to And~II, and a similar mean metallicity,
the abundance distribution in And~II is significantly broader than it is
in And~I\@.  This is illustrated in Fig.\ \ref{AndI_II} where the 
abundance distributions of the two dSphs are directly compared.
Quantitatively, 
$\sigma_{int}$([Fe/H]) $\approx$ 0.21 for And~I while it is 0.36 dex for 
And~II\@.  Similarly, the inter-quartile range is 0.36 dex versus 0.66 for
And~II, and
the central two-thirds of the sample range is 0.49 dex in And~I compared
to 0.80 dex for And~II\@.
{\it Thus, despite resulting in similar mean abundances, the chemical enrichment
processes in these two dSphs must have been significantly different}.  
The less luminous M31 dSph And~III also shows an intrinsic abundance dispersion
that is considerably less than that for And~II; Armandroff \etal (1993)
give 0.16 $\leq$ $\sigma_{int}$([Fe/H]) $\leq$ 0.24 from their ground-based
$V,I$ photometry.  As for Galactic dSphs, based on the compilation of Mateo
(1998) and the references therein, only the Fornax and Sagittarius dSphs
have intrinsic abundance dispersions that convincingly exceed that of 
And~II\@.  Both these dSphs, however, are at least two magnitudes more 
luminous than And~II.

The And~II abundance distribution shown in Fig.\ \ref{AndI_II} is sufficiently
broad compared to the abundance uncertainties that it is worth investigating
the implications of the observations through the use of simple chemical 
enrichment 
models (cf.\ \markcite{RZ80}Zinn 1978, \markcite{NF96}Norris \etal 1996).  
In particular, this process may shed some light on the hint of bimodality
seen in Fig.\ \ref{AndI_II} (cf.\ Fig.\ \ref{clr_sprd_fig}).  Of course in
reality the chemical evolution of a dSph galaxy is a complex process involving 
the effect of star formation driven processes, particularly supernovae, on the
distribution and chemical composition of the gas that resides within the
dark matter potential of the dSph.  Detailed models of these processes (e.g.\
\markcite{RL74}Larson 1974; \markcite{DS86}Dekel \& Silk 1986;
\markcite{PV86}Vader 1986, \markcite{PV87}1987; \markcite{MF99}Mac~Low \& 
Ferrara 1999) are,
however, necessarily heuristic in nature and therefore do not readily supply
an interpretative guide for individual objects in the way simple models do.

The simplest models are those in which the assumption of instantaneous
recycling is made (e.g.\ \markcite{BT80}Tinsley 1980).  The metallicity
distribution function $f(z)$ can be then written as:
\begin{equation}
f(z)\; =\; (1/y)\,exp\,[-(z-z_{0})/y]\;\;\;\;\;\;   z \geq z_{0} \label{eq1}
\end{equation}
Here $z_{0}$ is the initial abundance and $y$ is the ``yield''.  The 
relationship between the yield and the mean abundance then depends on 
how the gas is lost from the system.  If the rate of gas
loss is assumed to be directly proportional to the star formation rate 
(``steady gas loss'', e.g.\ \markcite{DH76}Hartwick 1976), then the yield
and the mean abundance, $<\!z\!>$, have the simple relation 
$y$ = $<\!z\!>$ -- $z_{0}$\@.  The abundance distribution for 
such simple models then has two parameters, $<\!z\!>$ and $z_{0}$.  
An alternative (cf.\ Zinn 1978) is ``sudden gas loss'' where the evolution is
terminated when the abundance reaches $z$ = $z_{max}$.  In this case the
mean abundance $<\!z\!>$ is again related to $y$, $z_{0}$ and $z_{max}$ and
we can employ equation (\ref{eq1}) for $z_{0}$ $\leq$ $z$ $\leq$ $z_{max}$ with
$f(z)$ = 0 otherwise.  The abundance distribution for these models then has
three parameters; $z_{0}$, $z_{max}$ and $y$ (or $<\!z\!>$).

\begin{figure}[p]
\plotfiddle{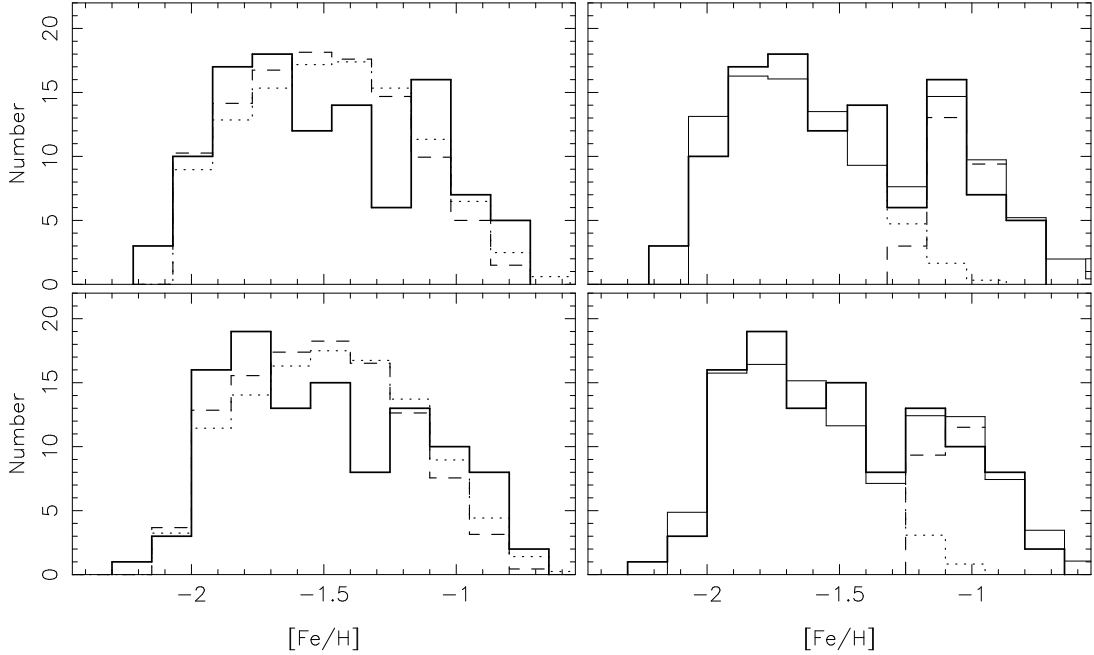}{4.5in}{-90.}{60.}{60.}{-280.}{350.}
\caption[fig10.ps]{The abundance distribution in And~II compared
to simple chemical enrichment models.  In all
panels the thick solid line is the observed distribution
inferred from the colors of the 108 red giants with 22.15 $\le$ $V$ $\le$ 
23.35.  The bin width is 0.15 dex.  The upper and lower panels show 
the histograms that result from offsetting the origin of the bins by 0.07 dex.
The left panels also show the predictions of one component ``steady gas loss''
(dotted line) and ``sudden gas loss'' (dashed line) models.
The right panels show the predictions of a two component steady gas loss
model.  Here the thin solid line is the sum of the two components.  The
dotted and the dashed lines show the contributions of the 
individual components in the region where they overlap.  
See Sect.\ \ref{abund_spd} for details of the parameters for this model.  
\label{feh_dist} }
\end{figure}

The left panels of Fig.\ \ref{feh_dist} show examples of these types
of models compared with the And~II observations.  In each case the
distribution function $f(z)$ was integrated over the same set of abundance bins
as the observations and then normalized to the total number of stars.  While
we have not thought it worthwhile to carry out any parameter optimization 
via, for example, a maximum likelihood analysis, the models shown are likely
to be representative of the ``best fit'' that can be achieved with models
of this type.  The steady gas loss model shown is for
log($<\!z\!>$/$z_{sun}$) = --1.35 and log($z_{0}$/$z_{sun}$) = --2.05 where
we have adopted $z_{sun}$ = 0.02.  For the sudden gas loss model
log($<\!z\!>$/$z_{sun}$) = --1.40, log($z_{0}$/$z_{sun}$) = --2.05 and
log($z_{max}$/$z_{sun}$) = --0.75. 

While these simple models do show reasonable agreement with the observations, 
the actual fits are not very satisfactory.  The models tend to have rather too
many stars at intermediate abundances and rather too few at the extremes 
compared to the observed distribution.  We have therefore investigated a
simple extension of these models (cf.\ Norris \etal 1996) in which we have
allowed for the possibility of {\it two} components, both following the
steady gas loss formalism.  Such a combination has five parameters:
$z_{0}$ and $<\!z\!>$ for each component and the relative strength of the two
components.  Again we have not attempted to find the optimum fit but show
in the right panels of Fig.\ \ref{feh_dist} an illustrative example which
certainly provides better agreement with the observations than the
single component models.  This model has 
log($<\!z\!>$/$z_{sun}$) = --1.6 and log($z_{0}$/$z_{sun}$) = --2.05 for
the ``metal-poor'' component and
log($<\!z\!>$/$z_{sun}$) = --0.95 and log($z_{0}$/$z_{sun}$) = --1.20
for the ``metal-rich'' component with metal-poor component stars outnumbering
metal-rich ones by 2.3 to 1.  We note also that a two component sudden gas 
loss model gives an almost identical representation of the data using
similar values for the mean abundances of the two components and their number 
ratio.

We therefore conclude that the broad abundance distribution in And~II, at
least in the context of simple enrichment models, is best understood in
terms of {\it two} components each with an intrinsic abundance range.
The mean abundances of these components differ by $\sim$0.6
dex and their number ratio is approximately 2.3 to 1 (metal-poor
to metal-rich).  This situation is reminiscent of that in the
Galactic globular cluster $\omega$ Cen, where Norris \etal (1996) find that
the [Ca/H] distribution among the cluster red giants suggests the presence
of two components whose mean [Ca/H] values differ by $\sim$0.5 dex.  It is 
also interesting to note that recently
\markcite{SM99}Majewski \etal (1999) have suggested that the 
metallicity distribution function in the Galactic dSph Sculptor is 
bimodal, with the
two components differing in mean abundance by $\sim$0.8 dex.  Taken 
{\it in toto} then, these and other available results for dSph abundance distributions (e.g.\ \markcite{TS99}Smecker-Hane \etal 1999) serve
to remind us yet again of the complex and diverse evolutionary histories 
these purportedly simple systems have undergone.

\subsection{The Age(s?) of And II} \label{age_sect}

The existence of blue horizontal branch and RR~Lyrae variable stars in And~II
immediately testifies to the existence of an old (age $>$ 10 Gyr) population
in this dSph galaxy.  This is not a surprising result since, although the
relative proportion of old stars varies significantly from dSph to dSph,
all Local Group dSph galaxies studied in sufficient detail contain an old population\footnote{The one possible exception to this statement is the 
Galactic companion dSph Leo~I in which 70 -- 80\% of the star formation
activity occurred between $\sim$7 and $\sim$1 Gyr ago (e.g.\ Gallart \etal 
1999a).  The detailed modelling of Gallart \etal (1999b), however, does 
not unambiguously
rule out the existence of a $\sim$12 -- 15 Gyr old population in Leo~I,
though any such population must be small.}.  

The blue horizontal branch (HB) stars make up $\sim$20\% of the total And~II HB 
population (cf.\ Sect.\ \ref{hbmorph}), and we can take this percentage
as an approximate lower limit on the fractional
size of the old population in And~II\@.  Nevertheless, the bulk of the 
HB stars in And~II are red and thus potentially of younger age.  However, we 
have already seen (cf.\ Sect.\ \ref{abund_spd}) that And~II has a substantial
internal abundance spread.  Consequently, at least in principle, significant
numbers of red HB stars could be contributed by a relatively metal-rich
old population, lessening the need for any younger stars.  Thus we need
to investigate whether the And~II HB morphology can be explained
by an appropriate combination of standard Galactic globular cluster HB
morphologies, given the consequent implication that 
most And~II stars would have ages similar to those of the Galactic globular 
clusters. 

\subsubsection{Modelling the Horizontal Branch Morphology}

To approach this question we have chosen to use published data for Galactic
globular clusters with known (old) ages rather than employ sets of
theoretical isochrones.  This is principally because theoretical isochrones
(at fixed age and abundance) do not reproduce the spread in color
(temperature) on the horizontal branch in real globular clusters without
invoking variable amounts of mass loss from the red giant branch progenitors.
On the other hand, to be of use in this simulation, where we will combine
appropriately scaled (in star number) c-m diagrams of globular clusters
of differing metal abundance, the cluster data needs to be essentially complete
to magnitudes fainter than that of the horizontal branch.  We have used the
photometry of M55 ([Fe/H] = --1.82) from \markcite{SL55}Lee (1977b), of
NGC~1851 ([Fe/H] = --1.29) from \markcite{AW92}Walker (1992; specifically
the data for his Fig.\ 4) and of 47 Tuc ([Fe/H] = --0.71) from
\markcite{SL47}Lee (1977a; specifically the data for his Ring 6). 
For all three of these cluster data sets the 
ratio of the number of bright red giants to the number of horizontal branch 
stars is virtually the same, indicating similar completeness levels.
Likely field objects that lie far from the fiducial sequences have been
removed as have known RR~Lyrae stars from the M55 and NGC~1851 
photometry.  

M55 is a ``standard'' metal-poor globular cluster with a strong blue HB
morphology, while 47 Tuc is the archetypal relatively metal-rich globular
cluster with a pure red HB morphology.  NGC~1851 has a somewhat anomalous
HB morphology in that the distribution of HB stars is approximately bimodal:
there are significant numbers of red and blue HB stars but a relative
deficiency of RR~Lyrae variables (see Fig.\ 7 of Walker 1992).  However,
the ages of all three clusters, based on main sequence turnoff photometry,
appear to be typical in the sense that they
are not notably younger (or older) than other clusters with similar abundance.
For example, none of these three clusters are distinguished in Fig.\ 1 of
\markcite{CD96}Chaboyer \etal (1996).

To combine these c-m diagrams we first shifted the observed cluster data
to our adopted And~II reddening and distance modulus.  We then split the
And~II abundance distribution of Fig.\ \ref{AndI_II} into three groups:
--2.2 $\leq$ [Fe/H] $\leq$ --1.6 to be represented by M55, 
--1.6 $\leq$ [Fe/H] $\leq$ --1.0 to be represented by NGC~1851, and
--1.0 $\leq$ [Fe/H] $\leq$ --0.5 to be represented by 47~Tuc.  These 
groups contribute 44, 45 and 11 percent of the total.  The numbers of 
stars in the M55 and 47~Tuc samples were then scaled so that the numbers
of bright red giants (we adopted $V$ $\leq$ 23.5) for these clusters
were in the correct proportion to the bright red giant numbers in the 
NGC~1851 sample.  The resulting c-m diagram is shown in the upper panel
of Fig.\ \ref{gcl_cmd_fig}.  Converting the $F450W-F555W$ color limits
for the HB morphology index $i$ (cf.\ Sect.\ \ref{hbmorph}) to $B-V$, we 
find $i$ = 0.46 $\pm$ 0.03 for this composite globular cluster sample.
For And~II, using the sample excluding the candidate variable stars
(i.e.\ Fig.\ \ref{bv_cmd}), the value of the index is $i$ = 0.17 $\pm$ 0.02,
a significantly lower value ($T_{2}$ = 8.3).  We therefore conclude that
the HB morphology of And~II cannot be modelled by the appropriate
combination of HB morphologies of ``standard'' globular clusters.

To reproduce the And~II HB morphology index, it is necessary to reduce the
numbers of blue HB stars in the upper panel of Fig.\ \ref{gcl_cmd_fig} and
to increase the numbers of red HB
stars.  To do this we have removed all of the NGC~1851 blue HB stars
and $\sim$45\% of the M55 blue HB stars and replaced them with red HB stars. 
These red HB stars were taken from the NGC~362\footnote{NGC~362 is well known
to be a ``second parameter'' cluster in that its HB morphology is redder than 
expected for its abundance.  Despite the continuing controversy surrounding 
the interpretation of the second parameter for Galactic globular clusters, 
it does seem that in the case of NGC~362 at least, the red HB morphology is 
caused by a younger age (e.g.\ \markcite{GN90}Green \& Norris 1990).} 
([Fe/H] = --1.28) c-m diagram of 
\markcite{WH82}Harris (1982, all listed red HB stars) and, since there
were insufficient stars in the NGC~362 
c-m diagram to make up the required
numbers, from the 47~Tuc data for the remaining $\lesssim$20 percent.
No changes were made to any non-HB stars.  The c-m diagram
with this modified HB morphology is shown in the lower part of 
Fig.\ \ref{gcl_cmd_fig}.  It has a HB morphology index $i$ value identical
to that for And~II\@.  

Thus to summarize this analysis, we find that in order to reproduce the
HB morphology of And~II, we have to convert $\sim$65\% of the total
HB population generated from an appropriate combination of ``standard'' 
globular clusters from blue into red HB stars.  Consequently, like many of the 
Galactic dSphs and like And~I (Paper~I) we can classify And~II as a ``second 
parameter'' object.  

The interpretation of this second parameter remains controversial, at least
for Galactic globular clusters.  On the one hand, \markcite{PS99}Stetson \etal 
(1999) have shown that the outer Galactic halo second parameter clusters
Pal~3, Pal~4 and Eridanus are indeed younger than their standard
globular cluster counterparts M3 and M5.  Yet on the other hand, recent results
for globular clusters in the Fornax dSph (\markcite{RB98}Buonanno \etal 1998)
and in the LMC (see the discussion in \markcite{GD99}Da~Costa 1999), for
example, appear to indicate
that age is not the only (second) parameter in determining HB morphologies.
However, for dSphs the interpretation of second parameter HB
morphologies as the result of younger-aged populations is less controversial.
There are at least two reasons for this.  First, dSphs are very low 
density systems
when compared to globular clusters.  Hence any dynamical effects that could
generate HB morphology changes will be insignificant.  Second,
among the Galactic dSphs, the second parameter nature of the horizontal
branch morphologies has, in most instances (e.g.\ Carina, Fornax, Leo~I 
and Leo~II),
been shown by main sequence turnoff photometry to result from significant
stellar populations that are younger (and in some systems substantially
younger) than Galactic halo globular clusters.  

There are, however, two
apparent exceptions to this ``age is the second parameter for dSphs''
interpretation.  The first of these is the low luminosity, metal-poor
Galactic dSph Draco.  \markcite{CG98}Grillmair \etal (1998) find from main
sequence turnoff photometry obtained with WFPC2 that the Draco
stellar population in the small field imaged is 1.6 $\pm$ 2.5 Gyr older
than the Galactic halo globular clusters M68 and M92.  Modulo possible
calibration uncertainties (cf.\ Sect.\ 3.2 of Grillmair \etal 1998), this
result contradicts, at the $\sim$1.4$\sigma$ level, the
expectation based on the respective HB morphologies and a strict ``age
is the second parameter'' interpretation, that Draco should be $\sim$2 Gyr
younger than these globular clusters.  The second exception is the
Galactic dSph Sculptor where the new ground-based photometry of
\markcite{HK99}Hurley-Keller \etal (1999) demonstrates vividly the 
HB morphology gradient in this dSph, previously discussed in
Paper~I\@, in which the red HB stars are more centrally concentrated than
the blue HB stars.  
However, at a 1--2 Gyr level of precision, Hurley-Keller \etal (1999) do not
see the corresponding $\sim$2 Gyr change in the mean age of the population 
that is required if age was solely responsible for the change in HB
morphology.

Despite these two possible exceptions, by analogy with the majority of
Galactic dSphs, we interpret our results for the
HB morphology in And~II as indicating that at least $\sim$50\% of the 
total population in this dSph has an age or ages less than that of 
the Galactic 
globular clusters.  Is there any independent evidence to support this
conclusion?  In the case of And~II there is, since, as mentioned in the
Introduction, there are claims of stars in And~II with luminosities 
considerably above that of the first giant branch tip.  Such upper-AGB stars
are likely to be of intermediate
(1 -- 10 Gyr) age.  We now look at those claims in more detail.  

\subsubsection{The Case for an Intermediate-Age Population} \label{ia_pop}

Aaronson \etal (1985) spectroscopically identified three candidate upper-AGB
stars in And~II\@.  These are the ``marginal carbon star'' A10, the M0 star
A209 and the carbon star A211.  Unfortunately A209 and A211 fell in the
vignetted region between the WF3 and WF4 frames and so we are unable to
include them in our c-m diagrams.  Similarly, A10 is outside the
WFPC2 field.  Nevertheless, we can calculate the bolometric magnitudes
for these stars using: (a) the original Aaronson \etal (1985) near-IR
photometry corrected
to our And~II modulus, and (b) the ($I$, $V-I$) photometry 
of C\^{o}t\'{e} \etal (1999b)
using bolometric corrections from \markcite{DA90}Da~Costa \& Armandroff (1990)
and our And~II distance modulus.  We can also utilize unpublished And~II
($I$, $V-I$) CCD photometry obtained with the 4~m telescope at Kitt Peak (see 
Armandroff 1994).  Considering first the most luminous star A209, which is
of type M rather than type C, the bolometric magnitude values are --4.3,
--4.8 and --4.4, respectively, from the above sources.  In particular,
the C\^{o}t\'{e} \etal (1999b) and the KPNO photometry differ by
$\Delta(V)$ = --0.23 and $\Delta(I)$ = --0.37 mag (sense is C\^{o}t\'{e} \etal
minus KPNO), so it is quite probable that star A209 is a variable, probably
of the long period type (LPV).  Such stars are not necessarily of 
intermediate-age.  As shown by \markcite{FE88}Frogel \& Elias (1988; see
also \markcite{GR97}Guarnieri \etal 1997 and the discussion in Sect.\ 7.2 of
\markcite{CA98}Caldwell \etal 1998), in Galactic globular clusters with
abundances [Fe/H] $\gtrsim$ --1.0 the Long Period Variables can have
luminosities as bright as M$_{bol}$ $\approx$ --4.6 (e.g.\ V7 in NGC~6712
at [Fe/H] = --1.01, or V2 in 47 Tuc at [Fe/H] = --0.71) yet there is no
suggestion that such clusters are of intermediate-age.  Since we have 
established that there is a population with [Fe/H] $\gtrsim$ --1.0 in 
And~II (cf.\ Fig.\ \ref{AndI_II}), there is no requirement to regard star
A209 as an intermediate-age object.

For the stars with carbon star spectral signatures, the M$_{bol}$ estimates 
range from --3.9 to --4.3 for A10 and suggest M$_{bol}$ $\approx$ --4.0 for
A211.  Again it is likely that both these stars are variable: in the same
sense as above $\Delta(V)$ = --0.07 and $\Delta(I)$ = --0.23 for A10 and
$\Delta(V)$ = +0.49 and $\Delta(I)$ = +0.12 mag for A211.  Nevertheless, the
bolometric magnitudes do place these stars above the RGB tip (M$_{bol}$ 
$\approx$ --3.6) and when coupled with the observed indications of third 
dredge-up material in their surface layers, argue convincingly
that they do belong to an intermediate-age population in And~II\@.  
In particular, all star clusters containing
stars showing evidence of the third dredge-up process (i.e.\ carbon
and S-type upper AGB stars) have main sequence turnoff ages less 
than $\sim$10 Gyr.
Similarly, those Galactic dSphs with upper-AGB carbon stars also have 
established main sequence populations of intermediate-age.  

The luminosities of the And~II carbon stars are, however, somewhat lower
than their counterparts in Galactic dSphs.  \markcite{MA94}Azzopardi (1994)
lists M$_{bol}$ $\approx$ --4.4, --4.5, --4.6 and --5.6 for the brightest
carbon stars in the Leo~II, Leo~I, Carina and Fornax dSph galaxies.  Of these
systems Leo~II contains the oldest intermediate-age population:
\markcite{MR96}Mighell \& Rich (1996) conclude from WFPC2 photometry of the
main sequence turnoff that it contains only stars older than $\sim$7 Gyr
with a ``typical'' member having an age of $\sim$9 Gyr.  While
two stars is far from sufficient to establish the AGB termination luminosity
in And~II (and thus the minimum age of the intermediate-age population) it 
seems likely that this age is nearer to $\sim$6--9 Gyr than it is to 
$\sim$1--3 Gyr.

C\^{o}t\'{e} \etal (1999b) list two additional candidate And~II intermediate-age
stars from their spectroscopic and photometric study: stars 14 and 30 in 
their list, both of which are radial velocity confirmed members.  
For star 14 the
C\^{o}t\'{e} \etal (1999b) ($I$, $V-I$) photometry yields M$_{bol}$ $\approx$
--4.5 on our adopted modulus but spectroscopically it is of type M 
(the spectrum reproduced
in Fig.\ 2 of C\^{o}t\'{e} \etal 1999b suggests a spectral type of M3 or 
M4~III).  Thus star 14 is similar to A209 and is not required to be of
intermediate-age.  The second object (star 30) has M$_{bol}$ $\approx$ --4.1 
and possesses a relatively featureless spectrum that rules out a C or M 
classification
(not surprising given the star's relatively blue $V-I$ color).  Unlike star 14
which lies some distance from the dSph's center, star 30 is within the field
imaged with WFPC2.  We find it to be 0.44 mag fainter in $V$ than do
C\^{o}t\'{e} \etal (1999b).  Furthermore, this star is one of the two at the
RGB tip in Fig.\ \ref{wfpc2cmd} that are classified as possible variables
(the two $F555W$ magnitudes differ by --0.09 mag).  Star 30 is then clearly
a variable which may have been near maximum light at the epoch of the 
C\^{o}t\'{e} \etal (1999b) photometry.  Again it is not necessary to classify
it as being of intermediate-age, though it could well be.  

Besides star 30, there are five other stars in common between our data set
and the spectroscopically observed sample of C\^{o}t\'{e} \etal (1999b): stars
3, 4, 5, 6 and 32.  
The $V$ photometry of C\^{o}t\'{e} \etal (1999b) agrees reasonably well with 
our WFPC2 data for stars 4 and 6 where the $\Delta$($V$) values, in the
sense (C\^{o}t\'{e} \etal -- WFPC2) are --0.08 and --0.04 mag, respectively.
For star 32 $\Delta$($V$) = --0.21, but inspection of the WFPC2 frames suggests
that the ground-based photometry could have been influenced by 
a nearby companion.
Star 5 is the second of the RGB tip candidate variables flagged in 
Fig.\ \ref{wfpc2cmd} where the two $F555W$ magnitudes differ by +0.08 mag.
However, the C\^{o}t\'{e} \etal (1999b) $V$ magnitude and the mean of the WFPC2 
values differ only by --0.07 mag (same sense as before) so that the
case for this star being a variable is not strong.  The final star in 
common is star 3, which C\^{o}t\'{e} \etal (1999b) have shown is a carbon
star with strong C$_{2}$ bands in its spectrum.  Our $V$ magnitude 
is 0.20 mag fainter than that given by C\^{o}t\'{e} \etal (1999b) and
thus this star is also likely to be a variable.  However, despite its carbon 
star type spectrum, its luminosity (M$_{bol}$ $\approx$ --3.45 from the 
C\^{o}t\'{e} \etal photometry) lies below that of the RGB tip and thus once
again it is not necessary to classify this star as an intermediate-age
object (C\^{o}t\'{e} \etal 1999b).  Near-infrared photometry 
would be worthwhile to confirm the relatively faint bolometric magnitude.

To summarize then, the case for an intermediate-age population in And~II
rests on the second parameter nature of at least half the horizontal
branch population and on the above-the-RGB-tip luminosity of the 
Aaronson \etal (1985) carbon and marginal-carbon stars A211 and A10.
An extensive survey for additional upper-AGB carbon stars in And~II is needed
to progress the question further. 

\subsubsection{The Nature of the Faint Blue Stars} \label{bss_age}

\begin{figure}[p]
\plotone{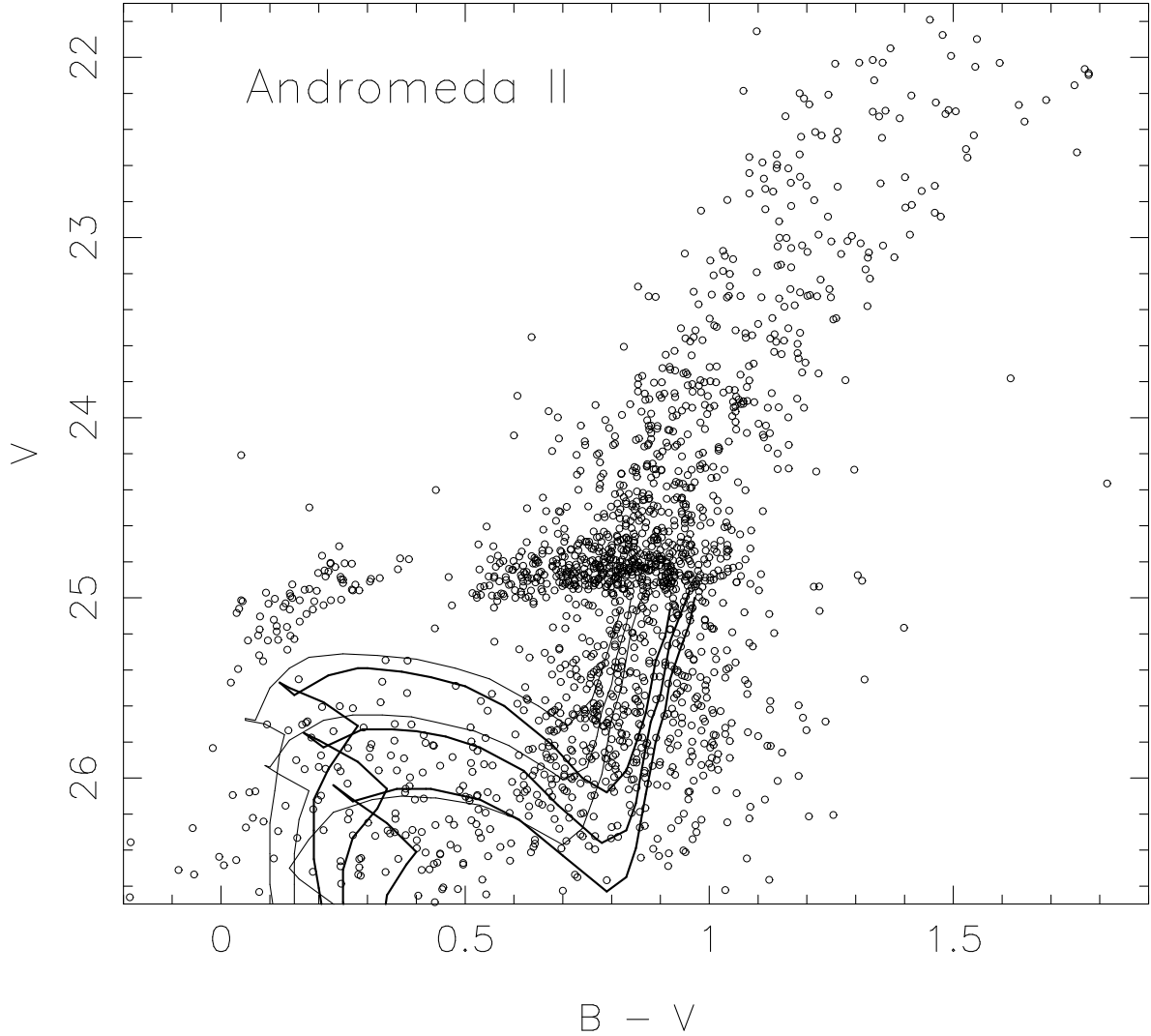}
\caption[fig12.ps]{A comparison of the And~II ``faint blue stars''
with theoretical isochrones from Bertelli \etal (1994).  The thick
lines are for an abundance log($z$/$z_{sun}$) = --0.7 and ages of 1.25,
1.6 and 2 Gyr.  The thin lines are for log($z$/$z_{sun}$) = --1.3 and ages
of 1.6, 2 and 2.5 Gyr.  Note that the data are seriously incomplete for
$V$ $\gtrsim$ 26.0 and $B$ $\gtrsim$ 26.7 mag.  \label{fbs_iso_fig}  }
\end{figure}

We now consider the faint blue stars discussed in Sect.\ \ref{fbs_sect}.
In Fig.\ \ref{fbs_iso_fig} isochrones from 
\markcite{BB94}Bertelli \etal (1994) are shown
superposed on the And~II observations
using our adopted reddening and distance modulus.  The isochrones are
for abundances log($z$/$z_{sun}$) = --1.3 and --0.7 dex, and for ages 
between 1.25 and 2.5 Gyr.  This figure shows that the location of these
faint blue stars in the And~II c-m diagram is consistent with the 
interpretation that they represent the tip of a main sequence population
whose youngest age is of order 1.5 Gyr.  Although we have not carried out
any quantitative completeness analysis, it is likely that the photometry
presented in Fig.\ \ref{fbs_iso_fig} is becoming seriously incomplete
by $V$ $\approx$ 26.0 (and $B$ $\approx$ 26.7).
Thus it is quite possible that there are substantially more of these stars
at $V$ $\gtrsim$ 26.0 mag.  Such stars would be either unevolved main sequence
stars of age $\sim$1.5 Gyr or a main sequence turnoff population that is
somewhat older.

Is this ``younger main sequence'' interpretation of these faint blue 
stars reasonable?  Perhaps, though
the apparent lack of upper-AGB carbon stars with M$_{bol}$ $\approx$ --4.5
or brighter suggests that any $\sim$1.5 -- 2.5 Gyr population cannot be a
large contributor to And~II's overall make-up.
Nevertheless, it is intriguing that we see relatively more of these faint
blue stars in And~II than in And~I (cf.\ Sect.\ \ref{fbs_sect}) and it is 
And~II where there are definite indications of an intermediate-age population
that is lacking in And~I\@.  We
note for completeness that for a population older than $\sim$6 Gyr, which
might provide the progenitors for the observed And~II upper-AGB carbon stars, 
the turnoff magnitude is fainter than $V$ $\approx$ 27.7, well below the 
detection threshold of the current data.

An alternative interpretation of these faint blue stars (cf.\ Paper~I) is
that they represent a blue straggler sequence arising from a postulated
age $\gtrsim$ 6 Gyr population.  Blue stragglers are known to occur in low
central concentration globular clusters so there seems no {\it a priori}
reason why they should not also be found in dSph galaxies.  Indeed, if blue
stragglers are related to binaries, as seems likely, then the high binary
star frequency inferred for the Draco and Ursa Minor Galactic dSphs by 
\markcite{OA96}Olszewski \etal (1996), which is considerably higher than that
for both Population~I and Population~II samples, would almost require the 
existence of blue straggler stars in dSph galaxies.  Further, 
Hurley-Keller \etal (1999) 
have offered a high binary fraction as the most reasonable explanation of
an unusual ``spur'' of stars observed in their c-m diagram for
the Sculptor dSph.
At $V$ $\approx$ 25.8 or M$_{V}$ $\approx$ +1.4, the brightest of the faint
blue stars in And~II are somewhat brighter than the brightest of the 
Galactic globular cluster blue straggler stars, but as noted in Paper~I
this is probably not a serious concern.  If there is a prominent And~II
population with an age less than that of the Galactic globular clusters,
then the turnoff masses are correspondingly higher and this presumably permits
somewhat more massive, and thus more luminous, blue straggler stars.

Unfortunately, a decision between these two interpretations of the And~II
faint blue star population cannot be made without significantly 
deeper photometry, whose acquisition represents a challenging task indeed.

\section{Discussion}   \label{discuss}

With HST/WFPC2 data now available for two of M31's six known dSph companions,
and new ground-based data appearing for the others (e.g.\ 
Caldwell 1999 and references therein) we can begin to compare 
the properties of the M31 dSph system with those of the Galaxy's dSph 
satellite system in some detail, seeking to identify any differences that
could be ascribed to the different environments of the dSph systems
(cf.\ Armandroff 1994, Armandroff \& Da~Costa 1999). In particular, M31
and the Galaxy are generally regarded as being quite similar, with the
major difference being M31's higher luminosity.  However, 
\markcite{KF99A}Freeman (1999a) argues that the stellar halos and bulges 
of these two galaxies are, in fact, fundamentally different.  M31 has a 
relatively metal-rich bulge with 
an $r^{1/4}$ law light profile that dominates any metal-poor halo population
at all radii (as in giant ellipticals).  This bulge probably formed rapidly
via violent collapse or early merger (Freeman 1999a).  The Galaxy, on the 
other hand, has a relatively metal-poor halo with a light profile
$\rho_{\rm L}$ $\propto$ $r^{-3.5}$ in which accretion of small satellite 
systems has probably played a significant role, particularly in the early stages
of halo formation.  Further, the Galaxy's bulge is
small, boxy and probably bar-like; it may have formed from the disk via
dynamical instabilities (Freeman 1999a).  Thus, in the present context, 
M31 and the Galaxy can be said to have provided different ``parental'' 
environments for their dSph systems.

We note first then that, as shown in Caldwell (1999) for example, the Galactic 
and M31 dSphs follow essentially
identical relations between central surface brightness, absolute magnitude
and mean abundance.  This result is reinforced if (minor) adjustments are
made to the location of the And~I and And~II points in the diagrams of
Caldwell (1999) to account for the mean metal abundance determinations 
presented here (cf.\ Sect.\ \ref{abund}), and for the smaller distance modulus
found for And~II in Sect.\ \ref{dist_sect}.  

There are only three significant outliers in these relations, as depicted
in Figs.\ 3 and 4 of Caldwell (1999).  The most
blatant of these is the Galactic dSph Leo~I which has a notably higher
central surface brightness for its mean metal abundance or absolute magnitude
when compared to the relation followed by the other dSph galaxies.  This 
discrepancy may be related to Leo~I's dominant relatively young population
(e.g.\ Gallart \etal 1999a, b) although, since Leo~I is less of an outlier
in the M$_{V}$,~$<$[Fe/H]$>$ plane, simple fading does not remove the
discrepancy.  The
second outlier is the Fornax dSph, whose published mean metallicity of approximately --1.4 dex (\markcite{BU85}Buonanno \etal 1985, 
\markcite{BE95}Beauchamp \etal 1995) is too low by $\sim$0.2 dex 
for this dSph's absolute magnitude.  The significance of this deviation from
the mean relation is not high, but here again the explanation may lie 
with the stellar population.  Fornax has a substantial
intermediate-age population (e.g.\ \markcite{PS98}Stetson \etal 1998 and
references therein) and this will generate a bluer mean giant branch color
(for a given abundance),
which could then be incorrectly interpreted as a lower mean abundance
when compared with standard globular cluster giant branches.  A spectroscopic
mean abundance determination for Fornax would aid in resolving (or confirming)
this discrepancy.  The third outlying object is And~V\@.  For this dSph
the mean abundance ($<$[Fe/H]$>$ $\approx$ --1.5, no error listed) derived by
Armandroff \etal (1998) from their c-m diagram study is too high by $\sim$0.5
dex for the central surface brightness and absolute magnitude determined
by Caldwell (1999).   And~V is a Cycle 8 target for observing with
HST/WFPC2 and it will be interesting to see if the new data confirm the 
relatively high abundance found by Armandroff \etal (1998).  If it does,
And~V will become the first dSph to clearly deviate significantly from
the M$_{V}$, $<$[Fe/H]$>$ relation followed by the other galaxies.  
Nevertheless, despite these possible outliers, the similarity of the
relations between central surface brightness, absolute magnitude and mean 
abundance for the M31 and Galactic dSphs indicates that these relations 
are not strongly influenced by the properties of the parent galaxy.

As regards horizontal branch morphologies, seven of eight Galactic dSphs
can be classified as ``second parameter'' objects in that they possess
redder HB morphologies than most Galactic globular clusters of abundance
similar to the dSph means.  The sole exception is the Ursa Minor dSph
which has a blue HB morphology consistent with its low mean abundance (the
HB morphology and mean abundance of the Sagittarius dSph are not yet well
established, so it is not included here).  With the And~II results presented 
in this paper, we now have HB morphologies for two of the six M31 dSph 
companions.  Like the vast majority of the Galactic dSph companions, 
both And~I and
And~II can be classified as second parameter objects.  For the Galactic
dSphs the second parameter nature of the HB morphology is, with the possible
exception of Draco and Sculptor, established via main sequence turnoff
photometry to result from significant populations of stars with intermediate
($\sim$1 -- 10 Gyr) ages.  Indeed, as inferred from the numbers and
luminosities of these intermediate-age stars, the
Galactic dSphs show a remarkable diversity in the star formation histories
(e.g.\ recent reviews of Da~Costa 1998 and Grebel 1999).  

Our results for And~I and And~II indicate that this diversity of star
formation histories applies also to the M31 dSph system.  While both 
And~I and And~II are second parameter systems, and thus probably 
contain sizeable intermediate-age populations (albeit perhaps of different
mean age) they differ considerably in detail.  And~II contains relatively
more blue HB stars and RR Lyrae variables than And~I (Sect.\ \ref{hbmorph})
indicating a higher fraction of older stars.  Yet And~II also has a relatively
higher fraction of faint blue stars (Sect.\ \ref{fbs_sect}) which
conceivably represent the tip of the main sequence for a $\sim$2 Gyr old
population.  And~II also has an identified population of upper-AGB 
(intermediate-age) stars (Sect.\ \ref{ia_pop}) which And~I apparently lacks. 
Further, And~II possesses a much broader internal abundance spread compared 
with And~I (cf.\ Fig.\ \ref{AndI_II}) despite the similar mean abundances of 
these two galaxies.  
On the other hand, And~I possesses a radial gradient in 
its HB morphology (cf.\ Paper~I) which And~II does not.  Thus there seems 
little doubt that even on the basis of just these two objects, we can conclude 
that the diversity of star formation and chemical enrichment histories seen 
in the Galactic dSphs is also seen among the M31 dSphs.  In that sense
M31's and the Galaxy's dSph satellite systems are again similar.  

Perhaps the only systematic difference that exists between these two sets of 
dSph galaxies is that the high luminosity markers of intermediate-age
populations, namely upper-AGB carbon stars, appear to be relatively less 
frequent in the M31 dSph systems than in the Galactic dSph systems.  
Such stars are known in And~II but aren't 
found in And~I or And~III (e.g.\ Armandroff 1994) and the published c-m 
diagrams for And~V, And~VI and And~VII don't suggest the presence of many
candidate upper-AGB stars in these systems either.
A specific search for carbon stars is called for, particularly as regards
the newly identified M31 dSphs.  If, however, this
difference in frequency of upper-AGB stars stands up to a more detailed and
complete analysis, then it may indicate that there is an
``environmental effect'' governing the evolution of these galaxies,
particularly as regards the fraction of the stellar populations with ages
less than $\sim$5--6 Gyr.  Such an environmental effect would be in the
sense that conditions for retaining gas are less favorable for satellites
of galaxies with massive, rapidly formed bulges (e.g.\ M31),
than they are for the satellites of small bulge systems like the Milky Way.
Such an effect is not inconsistent with the conclusions of van den Bergh (1994,
see also \markcite{ES74}Einasto \etal 1974) who noted that ram pressure
stripping in gaseous coronae, supernova-driven galactic winds and a high UV
flux from a parent galaxy could all affect the evolution of a satellite
dwarf.  The satellites most affected would be those which are less massive or
less dense and which are on orbits that take them to smaller galactocentric
distances (e.g.\ \markcite{HK97}Hirashita \etal 1997).

One more point concerning the M31 and the Galactic dSph satellite systems
can be made.
As commented on by a number of authors (e.g.\ Armandroff \etal 1999, Grebel \& Guhathakurta 1999), with the addition of the newly discovered objects to
the list of M31 dSph companions, and now the confirmation here that And~II
is associated with M31 and not M33, it is apparent that the systems of
dSph galaxies associated with M31 and the Galaxy are roughly equivalent
in size.  For both galaxies the radius of the system of known dSph satellites
is $\sim$250 kpc.
The outermost members of these satellite systems are, of course,
of prime importance in determining the mass and extent of the dark matter
halos of the parent galaxies (cf.\ \markcite{ZO89}Zaritsky \etal 1989).  
With large ground-based telescopes it is now possible to measure line-of-sight 
velocities for the M31 dSph satellites
and thus use these objects to provide further constraints on M31's mass.
For example, \markcite{CM99}C\^{o}t\'{e} \etal (1999a) give a heliocentric
velocity of --188 $\pm$ 3 km s$^{-1}$ for And~II from 9 measurements of 7
red giant members.  Following \markcite{LB99}Lynden-Bell (1999) this velocity
corresponds to a line-of-sight velocity V$_{\ell}$ = +82 km s$^{-1}$ for 
And~II in a frame in which M31 is at rest.  Then, using 
M $\approx$ 3$<$V$_{\ell}^{2}>$R/G (cf.\ Lynden-Bell 1999) with our 
And~II -- M31 distance of $\sim$190 kpc (cf.\ Sect.\ \ref{dist_sect}), 
a mass estimate 
for M31 of 9~$\times$~10$^{11}$~M$_{\sun}$
results.  While obviously additional objects are needed to improve the 
precision of this estimate, it is at least consistent with existing M31 mass
determinations (e.g. 24~$\times$~10$^{11}$~M$_{\sun}$, 
\markcite{KF99B}Freeman 1999b
and references therein).  It is also consistent with the results of
\markcite{ZS93}Zaritsky \etal (1993, see also 
\markcite{ZW94}Zaritsky \& White 1994) who found masses of order 
20~$\times$~10$^{11}$~M$_{\sun}$ within radii of $\sim$200 kpc from an ensemble
analysis of satellite velocities for a set of isolated late-type spirals
with luminosities similar to that of the Galaxy.  Clearly line-of-sight
velocities for the other M31 dSph companions are going to play a vital role 
in constraining the mass of M31's extended dark halo.  In this context it is
worth noting that the observed line-of-sight velocities for M31 satellites
contain contributions from both the tangential and radial velocity components
of the satellite.  This contrasts with the situation for our Galaxy where the
line-of-sight velocities for distant satellites are dominated by the 
radial component.  Thus the M31 satellite velocities mass estimate will
be less sensitive to assumptions regarding the distribution of orbital
eccentricities than is the case for the Galaxy.

\acknowledgements
T.\ E.\ A.\ is grateful to Prof.\ Jeremy Mould, Director of the Research
School of Astronomy \& Astrophysics, ANU, for support of a sabbatical
visit during which some of the work reported here was carried out.  The
authors are also grateful to the NOAO WIYN Queue team, particularly Di Harmer
and Paul Smith, for their efforts in obtaining the WIYN images used in this
paper, and to Alistair Walker for supplying a copy of his NGC~1851 photometry.
The comments of John Norris on a draft of this manuscript were also
appreciated.  This research was supported in part
by NASA through grant number GO-06514 from the Space Telescope Science
Institute, which is operated by AURA, Inc., under NASA contract NAS 5-26555.

\newpage

\appendix

\section{A Revised Analysis of the Andromeda I Data}

\begin{figure}[p]
\plotfiddle{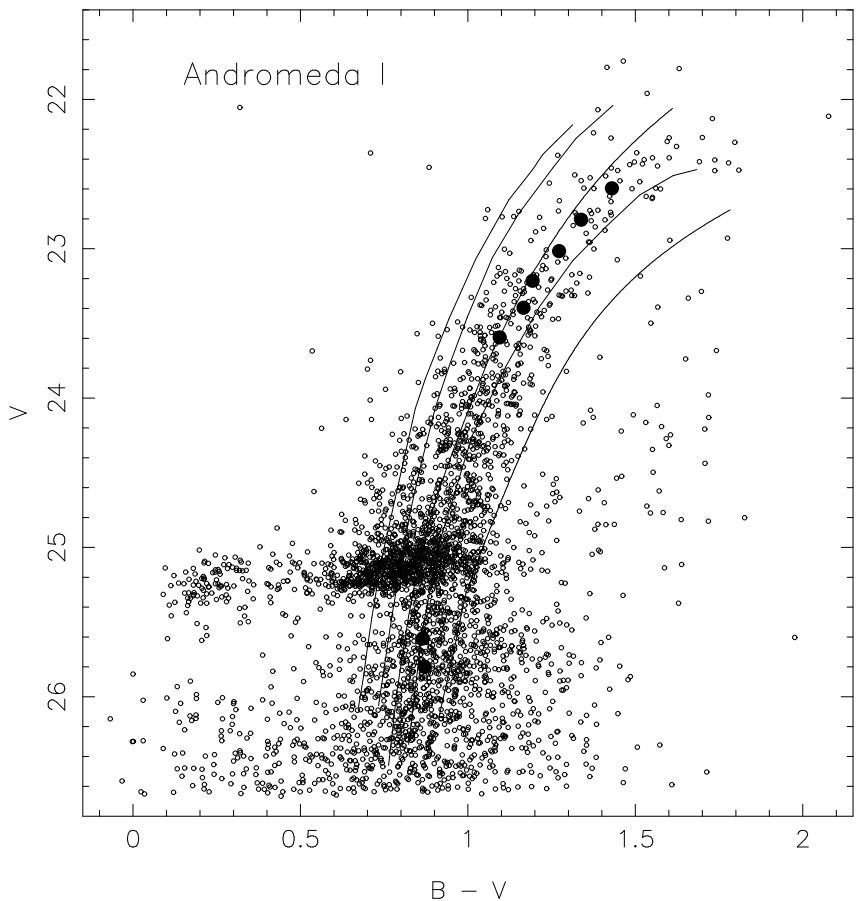}{5.3in}{0.}{70.}{70.}{-250.}{-145.}
\caption[fig13.ps]{A color-magnitude diagram for Andromeda~I based
on a reanalysis of the photometry presented in Paper~I\@.  The data shown are
for the three WF CCDs and no attempt has been made to exclude any candidate
variables.  The $B-V$ zeropoint has been shifted redwards by 0.055
mag relative to the transformation of Holtzman \etal (1995b), in accordance
with the discussion in Sect.\ \ref{bv_zpt}.  Shown superposed on the 
revised photometry are the giant branches of the standard
globular clusters M68 ([Fe/H]=--2.09), M55 ([Fe/H]=--1.82), NGC~6752 
([Fe/H]=--1.54), NGC~362 ([Fe/H]=--1.28) and 47~Tuc ([Fe/H]=--0.71).  An And~I
apparent distance modulus of (m -- M)$_{V}$ = 24.66 and a reddening of
E($B-V$) = 0.05 have been used to place the globular cluster giant branch
data in this diagram.  The filled symbols give the mean And~I giant branch 
colors for $\pm$0.1 $V$ mag bins.  \label{andi_bv_cmd} }
\end{figure}

In Paper~I we analyzed our HST/WFPC2 photometry of Andromeda~I using the
0.04 mag/800 pixel ramp correction for CTE effects, the Holtzman \etal
(1995b) zeropoints for the $F555W$ to $V$ and $F450W$ to $B$ transformations,
and the Burstein \& Heiles (1982) reddening estimate.  All three of
these processes or assumptions, however, have been superseded by new techniques
or new data.  Thus, to allow a proper comparison of
Andromeda~I with the Andromeda~II results presented in this paper, we have
reanalyzed the And~I photometry in a manner analogous to that used for
And~II\@.  In particular, we have applied the WH97 CTE correction formalism,
as described in Sect.\ \ref{cte_sect}, to the original And~I aperture 
photometry measurements.  We have also used
the adjusted $B-V$ zeropoint discussed in Sect.\ \ref{bv_zpt} and
restricted the data to just that for the WF CCDs.  The resulting And~I
c-m diagram is shown in Fig.\ \ref{andi_bv_cmd}\@.  Note that candidate 
variable stars in And~I have not been excluded from this c-m diagram 
(cf.\ Fig.\ \ref{bv_cmd}).

The majority of the And~I results presented in Paper~I are not changed by this
reanalysis.  In particular, the horizontal branch morphology, the horizontal 
branch morphology gradient, and the inferences drawn concerning the age of 
the bulk of the stellar population in And~I
remain unaffected by the small changes in the photometric data.  There are,
however, some minor modifications which we outline here.  First, in the revised
photometry, the mean magnitude of the horizontal branch is slightly brighter
than the value presented in Paper~I ($V$(HB) = 25.23 $\pm$ 0.04 rather
than 25.25 $\pm$ 0.04).  Further, the Schlegel \etal (1998) reddening maps
predict E($B-V$) = 0.053 for And~I; we adopt E($B-V$) = 0.05 $\pm$ 0.01
and A$_{V}$ = 0.17 $\pm$ 0.03 mag.  These values are slightly higher than
those adopted in Paper~I\@.  Consequently, for our And~I mean abundance of 
$<$[Fe/H]$>$ = --1.46 $\pm$ 0.12 (see following paragraph) and our
adopted distance scale, the line-of-sight distance to And~I is marginally
smaller than the value given in Paper~I\@.  We find 
(m--M)$_{0}$ = 24.49 $\pm$ 0.06 corresponding to a distance of 790 $\pm$ 25 
kpc.  The difference between the Paper~I value (810 $\pm$ 30 kpc) and the
new distance is sufficiently small, however, that the relative line-of-sight 
distance between M31 and And~I given in Paper~I, 
0 $\pm$ 70 kpc, is essentially unaltered.  This is
also the case for the true distance of And~I from the center of M31 -- it
lies between $\sim$45 and $\sim$85 kpc, as discussed in Paper~I.

As regards the mean metallicity of And~I, applying the method of analysis
described in Sect.\ \ref{abund} to the revised photometry yields a mean
abundance of $<$[Fe/H]$>$ = --1.46 $\pm$ 0.12 from six 0.2 $V$ mag wide bins
on the upper giant branch and $<$[Fe/H]$>$ = --1.20 $\pm$ 0.25 from two
similar bins on the lower giant branch.  Here the uncertainty given for each
abundance includes the
statistical error in the mean colors, the effects of a ($B-V$) zeropoint
uncertainty of $\pm$0.03 mag and a distance modulus uncertainty of $\pm$0.06
mag, and the effect of uncertainty in the abundance calibration derived 
from the standard globular cluster giant branches.  As for And~II no one
uncertainty dominates the total error for the upper giant branch value,
while for the lower giant branch, the ($B-V$) zeropoint uncertainty is the
major contributor to the abundance error.  It is worth noting that these
abundance estimates have been derived without applying any additional color
shift to the globular cluster standard giant branches, as was found necessary
in Paper~I\@.  A comparison of the standard globular giant branches with
the new And~I photometry for an apparent distance modulus of (m--M)$_{V}$ = 
24.66 and a reddening E($B-V$) = 0.05 mag is also shown in 
Fig.\ \ref{andi_bv_cmd}\@.  This figure also shows the mean 
giant branch colors for the adopted 0.2 $V$ mag wide bins
(cf.\ Fig.\ \ref{bv_cmd_gb}).

\begin{figure}[p]
\plotfiddle{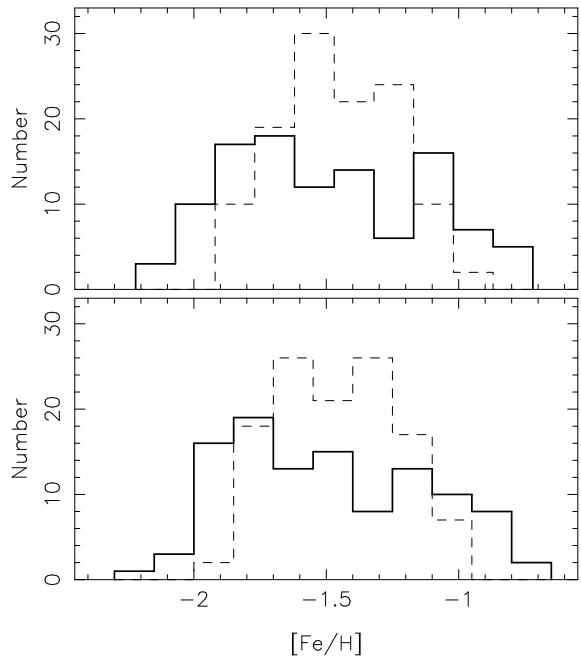}{4.5in}{-90.}{60.}{60.}{-300.}{300.}
\caption[fig9.ps]{A comparison of the abundance distributions for
the dSphs And~I (dashed line) and And~II (solid line).  These abundance
distributions have been derived in an identical manner from similar WFPC2
data.  The And~I sample contains 117 red giants while that for And~II
contains 108 stars.  The bin width is 0.15 dex. The two panels show the 
effect of offsetting the origin of the histogram bins by 0.07 dex.  
\label{AndI_II} }
\end{figure}

Because of the additional uncertainty in the lower giant branch mean colors
caused by contamination from (relatively metal-rich) M31 field
giants, and for the reasons discussed in Sect.\ \ref{abund} above,
we regard the And~I mean abundance derived from the upper giant branch
as the more reliable determination.  Surprisingly, despite the reanalysis
carried out here, this mean abundance, $<$[Fe/H]$>$ = --1.46 $\pm$ 0.12,
has turned out to be very similar to the value preferred in Paper~I
$<$[Fe/H]$>$ = --1.45 $\pm$ 0.2)\@.  This is also true of the intrinsic
abundance spread in And~I\@.  Applying the analysis technique described
in Sect.\ \ref{abund_spd} to the revised And~I photometry results in 
$\sigma_{int}$([Fe/H]) $\approx$ 0.21 dex, an inter-quartile range of
0.36 dex and a central two-thirds of the sample range of 0.49 dex.  The
corresponding values in Paper~I were 0.20, 0.30 and 0.45 dex, respectively.
The And~I abundance distribution from the 117 stars with 22.5 $\leq$ $V$
$\leq$ 23.5 is shown as the dashed line in Fig.\ \ref{AndI_II}.  This
distribution is clearly narrower than that of And~II and is approximately
gaussian in shape.

\newpage

\begin{figure}[p]
\caption{The location of the section of And~II
imaged with WFPC2 is shown superposed on a ground-based image of this
dSph obtained with the WIYN telescope.  The WIYN image is the combination 
of three 400 s exposures in the $V$ band and has images with FWHM $\approx$ 
0.55$\arcsec$.  North is at the top and East is to the left.
\label{wiynpic} }
\end{figure}

\begin{figure}
\caption{A mosaic of the 
And II WFPC2 field made from the
combination of four 1200~s $F555W$ exposures.  North is indicated by the direction 
of the arrow and East by the line.  Both indicators are 10$\arcsec$ in
length.  The center of And~II lies near the WF3/WF4 overlap region close 
to the edge of the frame.  The majority of WF2 then lies beyond And~II's
core radius.  This positioning was necessary to avoid bright foreground
stars.  \label{wfpc2pic} }
\end{figure}

\begin{figure}[p]
\plotone{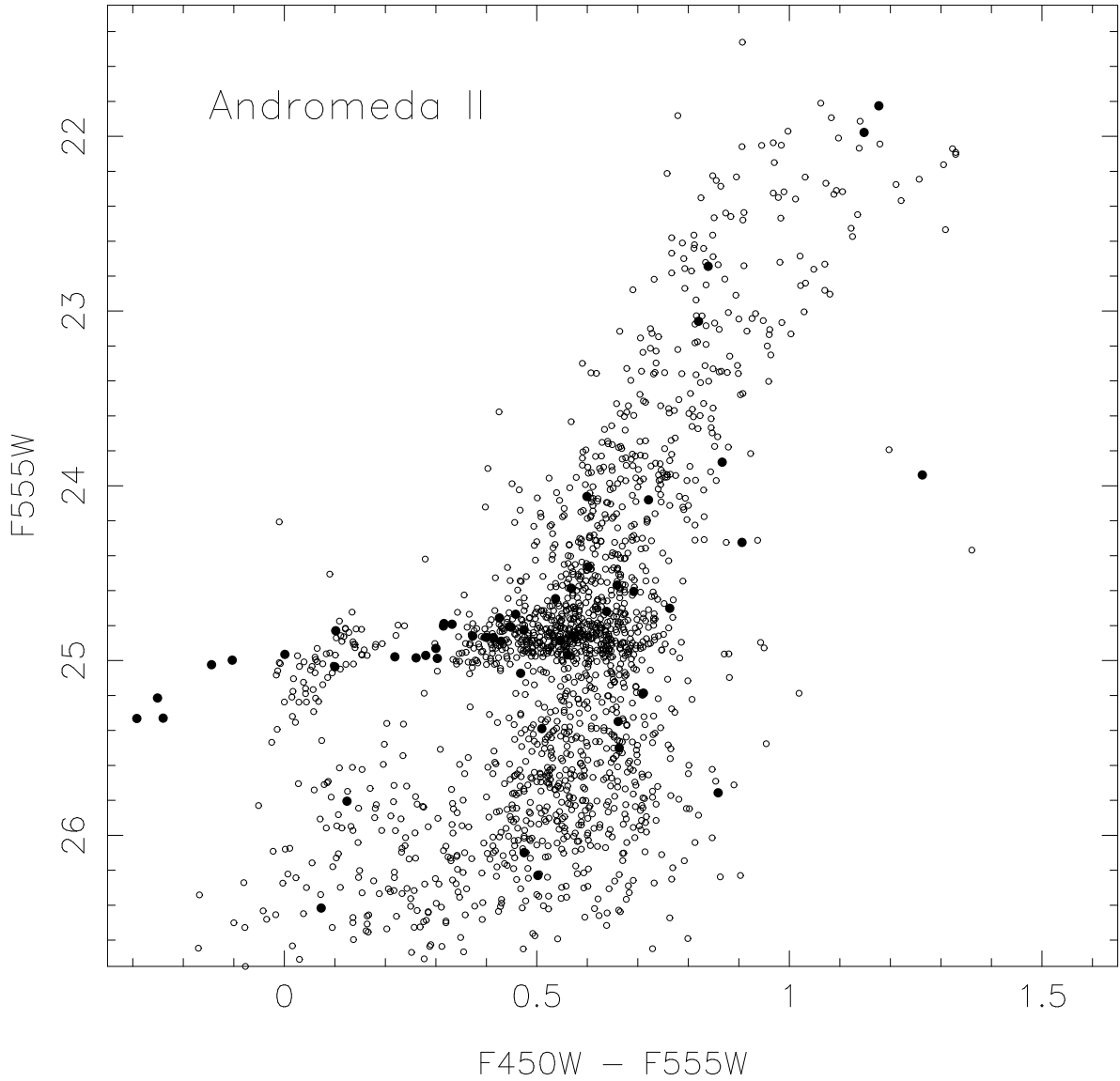}
\caption[fig3.ps]{The combined Andromeda~II color-magnitude diagram
on the WFPC2 system of Holtzman \etal (1995b) after the application of the Whitmore \& Heyer (1997)
CTE corrections.  Stars from all three WF CCDs are plotted.  The filled 
symbols are stars which show large differences between the two sets of 
observations.  The majority of these stars are likely
to be And~II variables.  \label{wfpc2cmd} }
\end{figure}

\begin{figure}[p]
\plotone{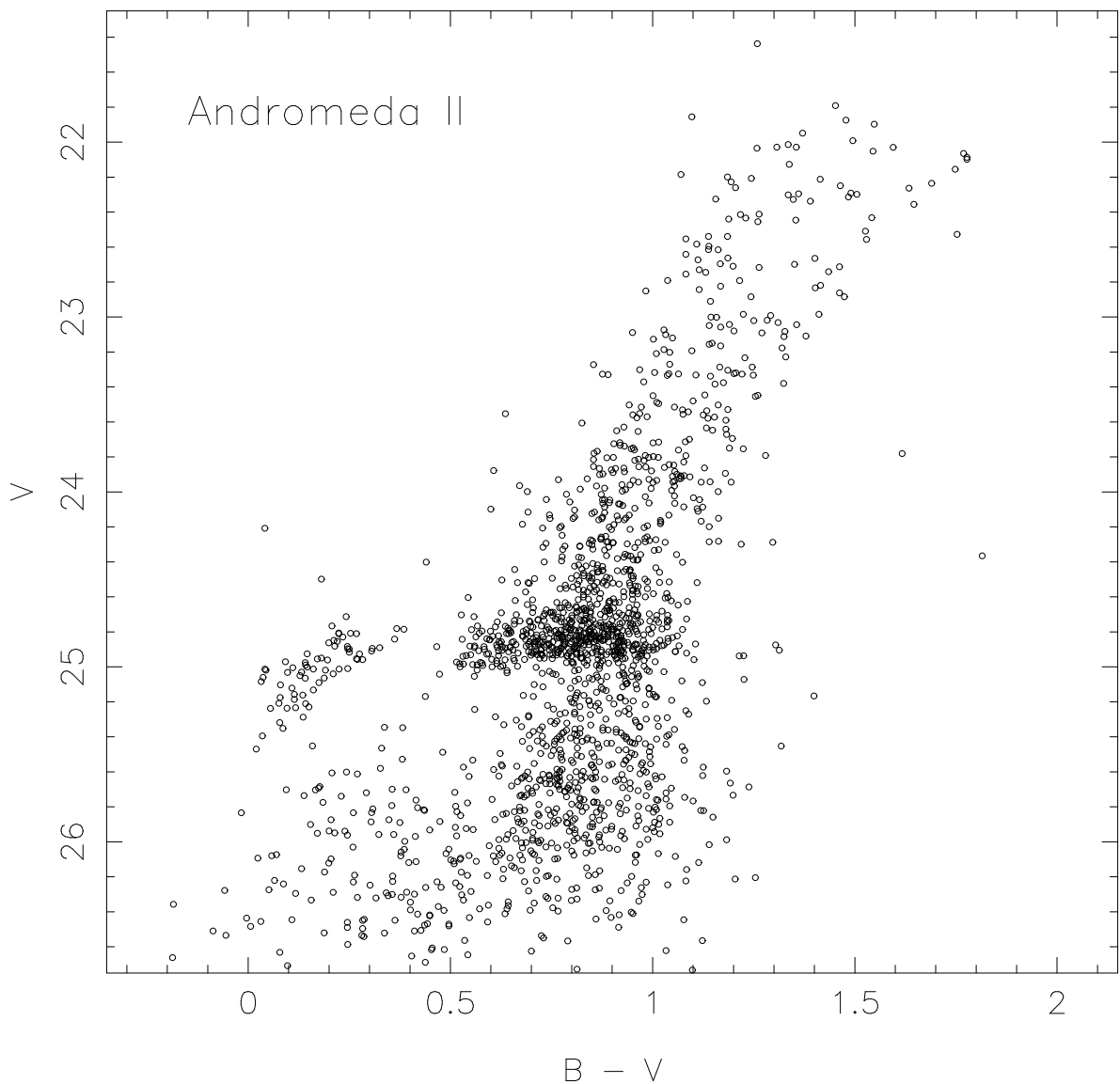}
\caption[fig5.ps]{The combined Andromeda~II color-magnitude diagram
on the standard $B$,$V$ system.  The $B-V$ zeropoint has been shifted 
redwards by 0.055
mag relative to the transformation of Holtzman \etal (1995b), in accordance
with our ground-based determination of the zeropoint (see Sect.\ \ref{bv_zpt}).
Candidate variables (cf.\ Fig.\ \ref{wfpc2cmd}) have been excluded from this 
plot.   \label{bv_cmd} }
\end{figure}

\begin{figure}[p]
\plotone{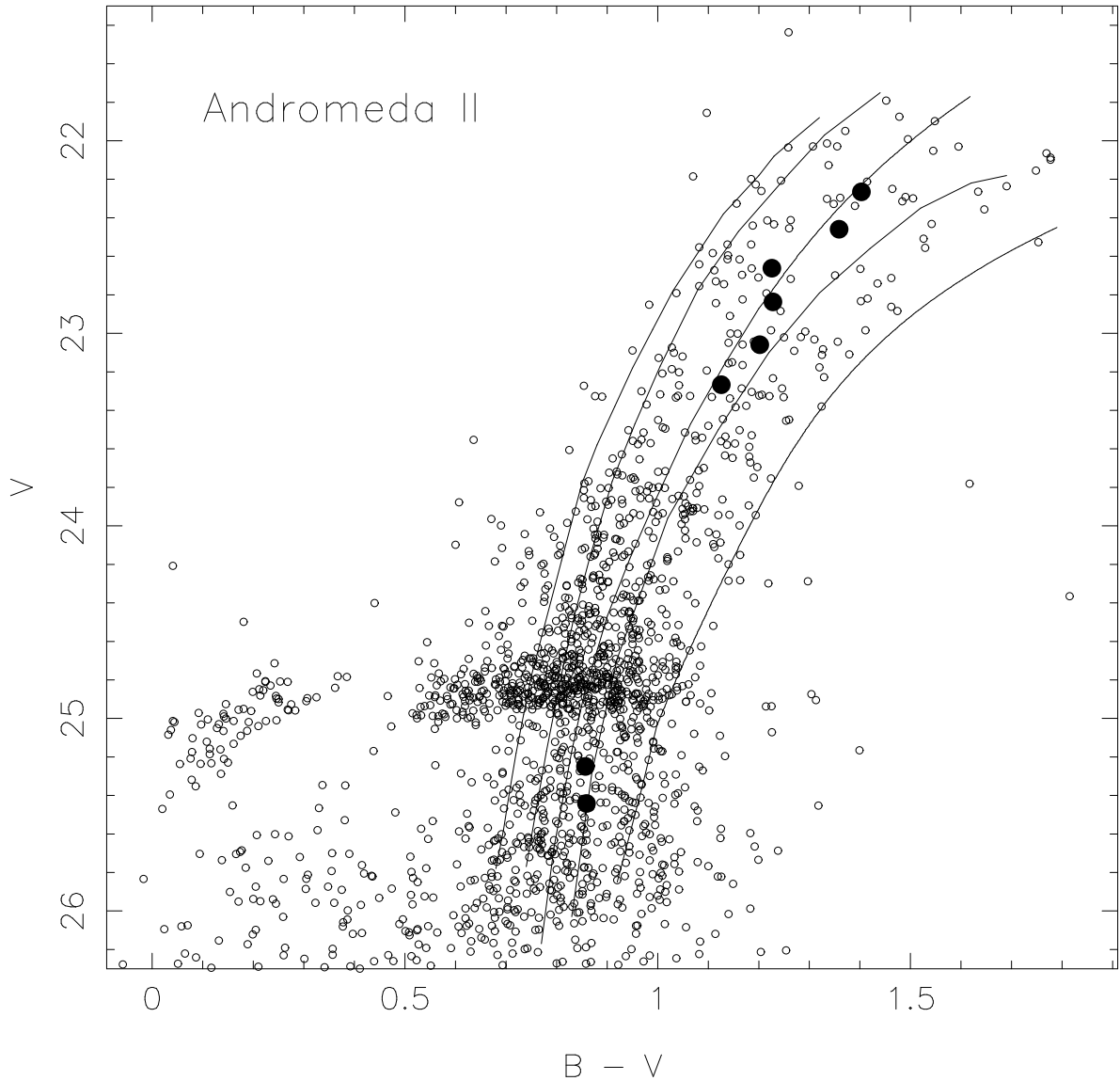}
\caption[fig8.ps]{The Andromeda~II ($V$, $B-V$) color-magnitude 
from 
Fig.\ \ref{bv_cmd} is shown superposed with the giant branches of the standard
globular clusters M68 ([Fe/H]=--2.09), M55 ([Fe/H]=--1.82), NGC~6752 
([Fe/H]=--1.54), NGC~362 ([Fe/H]=--1.28) and 47~Tuc ([Fe/H]=--0.71).  An And~II
apparent distance modulus of (m -- M)$_{V}$ = 24.36 and a reddening of
E($B-V$) = 0.06 have been used to place the globular cluster giant branch
data on this diagram.  The filled symbols give the mean And~II giant branch 
colors for $\pm$0.1 $V$ mag bins.  \label{bv_cmd_gb} } 
\end{figure}

\begin{figure}[p]
\plotfiddle{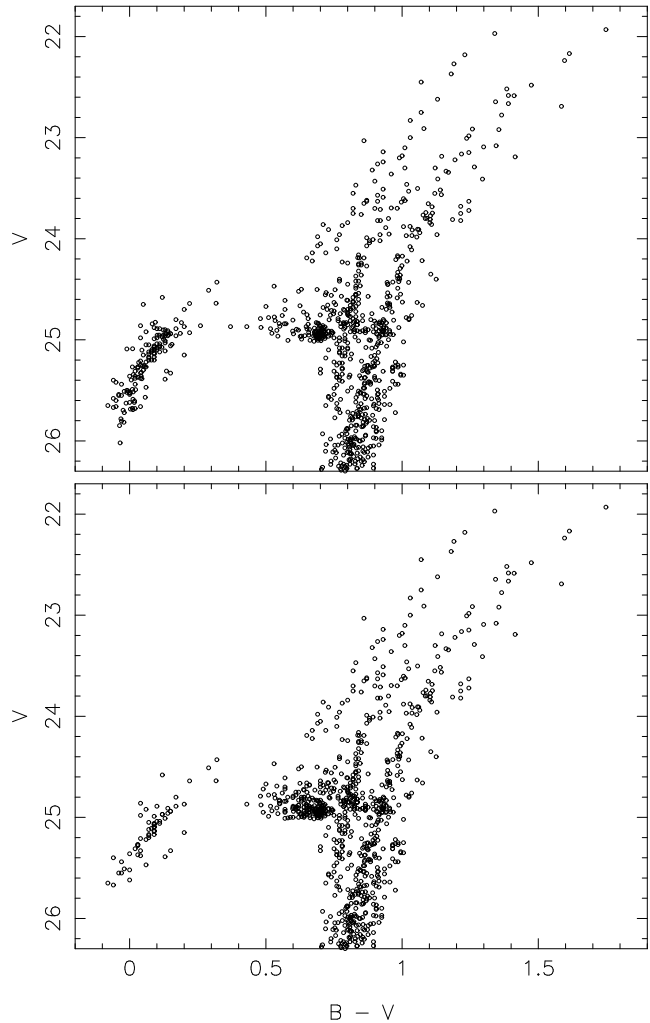}{4.5in}{0.}{60.}{60.}{-150.}{-45.}
\caption[fig11.ps]{The upper panel is a composite c-m diagram made
up from observed c-m diagrams for the Galactic globular clusters M55
([Fe/H] = --1.82), NGC~1851 ([Fe/H] = --1.29) and 47~Tuc ([Fe/H] = -0.71),
with the relative star numbers (44:45:11) scaled to reflect the And~II abundance
distribution.  This c-m diagram clearly has relatively more blue horizontal 
branch (HB) stars than does And II (cf.\ Fig.\ \ref{bv_cmd}).  In the lower 
panel all the NGC~1851 blue HB and 40\% of the M55 blue HB stars have been
replaced with red HB stars from NGC~362 ($\sim$80\%) and 47~Tuc ($\sim$20\%).
The horizontal branch morphology of this c-m diagram is now a better match
to that of And~II.   \label{gcl_cmd_fig} }
\end{figure}

\clearpage
\begin{deluxetable}{lcccc}
\tablecolumns{5}
\tablewidth{0pt}
\tablenum{1}
\tablecaption{Photometric Errors.}
\tablehead{\colhead{F555W}&\colhead{Mean Error}&\colhead{Mean Error in}
&\colhead{F450W}&\colhead{Mean Error}\nl
&\colhead{in F555W}&\colhead{F450W--F555W}&&\colhead{in F450W}}
\startdata
21.5--23.0 & 0.012 & 0.014 & 22.0--24.4 & 0.017\nl
23.0--24.0 & 0.014 & 0.022 & 24.4--24.9 & 0.023\nl
24.0--24.5 & 0.019 & 0.033 & 24.9--25.4 & 0.036\nl
24.5--25.0 & 0.032 & 0.046 & 25.4--25.8 & 0.044\nl
25.0--25.4 & 0.038 & 0.066 & 25.8--26.2 & 0.065\nl
25.4--25.8 & 0.050 & 0.077 & 26.2--26.6 & 0.086\nl
25.8--26.1 & 0.068 & 0.108 & 26.6--26.9 & 0.122\nl
26.1--26.4 & 0.077 & 0.136 & 26.9--27.2 & 0.151\nl

\enddata 
\label{error_tab}
\end{deluxetable}


\begin{references}

\reference{AM85} Aaronson, M., Gordon, G., Mould, J., Olszewski, E., \&
Suntzeff, N. 1985, \apj, 296, L7
\reference{TA89} Armandroff, T. E. 1989, \aj, 97, 375
\reference{TA94} Armandroff, T. E. 1994, in An ESO/OHP Workshop on Dwarf 
Galaxies, eds G. Meylan \& P. Prugniel, (Garching, ESO), p.\ 211
\reference{AD99} Armandroff, T. E., \& Da Costa, G. S. 1999, in The Stellar
Content of Local Group Galaxies, IAU Symposium 192, eds P. Whitelock \& R.
Cannon, (San Francisco, ASP), p.\ 203
\reference{AD93} Armandroff, T. E., Da Costa, G. S., Caldwell, N., 
\& Seitzer, P. 1993, \aj, 106, 986
\reference{TA98} Armandroff, T. E., Davies, J. E., \& Jacoby, G. H. 1998,
\aj, 116, 2287
\reference{AJ99} Armandroff, T. E., Jacoby, G. H., \& Davies, J. E. 1999,
\aj, 118, 1220
\reference{MA94} Azzopardi, M. 1994, in The Local Group: Comparative and
Global Properties, eds A. Layden, R. C. Smith, \& J. Storm, (Garching, ESO),
p.\ 129
\reference{BH39} Baade, W., \& Hubble, E. 1939, \pasp, 51, 40
\reference{BE95} Beauchamp, D., Hardy, E., Suntzeff, N. B., \& Zinn, R.
1995, \aj, 109, 1628
\reference{BB94} Bertelli, G., Bressan, A., Chiosi, C., Fagotto, F., \&
Nasi, E. 1994, \aaps, 106, 275
\reference{BU85} Buonanno, R., Corsi, C. E., Fusi Pecci, F., Hardy, E., \&
Zinn, R. 1985, \aap, 152, 65
\reference{RB98} Buonanno, R., Corsi, C. E., Zinn, R., Fusi Pecci, F., 
Hardy, E., \& Suntzeff, N. B. 1998, \apj, 501, L33
\reference{BH82} Burstein, D., \& Heiles, C. 1982, \aj, 87, 1165
\reference{NC99} Caldwell, N. 1999, \aj, 118, 1230
\reference{NC92} Caldwell, N., Armandroff, T. E., Seitzer, P., \& Da~Costa,
G. S. 1992, \aj, 103, 840
\reference{CA98} Caldwell, N., Armandroff, T. E., Da~Costa, G. S., \&
Seitzer, P. 1998, \aj, 115, 535 
\reference{CD96} Chaboyer, B., Demarque, P., \& Sarajedini, A.  1996, \apj,
459, 558
\reference{CM99} C\^{o}t\'{e}, P., Mateo, M., Olszewski, E. W., \& Cook, K. H.
1999a, \apj, in press
\reference{CO99} C\^{o}t\'{e}, P., Oke, J. B., \& Cohen, J. G. 1999b, \aj,
118, 1645
\reference{GD98} Da Costa, G. S. 1998, in Stellar Astrophysics for the Local 
Group, eds A. Aparicio, A. Herrero, \& F. S\'{a}nchez, (Cambridge, Cambridge 
Univ.\ Press), p.\ 351
\reference{GD99} Da Costa, G. S. 1999, in New Views of the Magellanic Clouds,
IAU Symposium 190, eds Y.-H. Chu, N. B. Suntzeff, J. E. Hesser \& D. Bohlender,
(San Francisco, ASP), in press
\reference{DA90} Da Costa, G. S., \& Armandroff, T. E. 1990, \aj, 100, 162
\reference{DA96} Da Costa, G. S., Armandroff, T. E., Caldwell, N., \& Seitzer,
P. 1996, \aj, 112, 2576 (Paper~I)
\reference{DS86} Dekel, A., \& Silk, J. 1986 \apj, 303, 39
\reference{ES74} Einasto, J., Saar, E., Kaasik, A., \& Chernin, A. D. 1974,
\nat, 252, 111
\reference{KF99A} Freeman, K. C. 1999a, in The Third Stromlo Symposium: The
Galactic Halo, ASP Conf.\ Ser.\ 165, eds B. K. Gibson, T. S. Axelrod, \& 
M. E. Putman, (San Francisco, ASP), p.\ 167
\reference{KF99B} Freeman, K. C. 1999b, in The Stellar Content of Local Group 
Galaxies, IAU Symposium 192, eds P. Whitelock \& R. Cannon, (San Francisco, 
ASP), p.\ 383
\reference{FE88} Frogel, J. A., \& Elias, J. H. 1988, \apj, 324, 823
\reference{FP96} Fusi Pecci, F., Buonanno, R., Cacciari, C., Corsi, C. E.,
Djorgovski, S. G., Federici, L., Ferraro, F. R., Parmeggiani, G., \& Rich, R. M.
1996, \aj, 112, 1461
\reference{G99A} Gallart, C., \etal 1999a, \apj, 514, 665
\reference{G99B} Gallart, C., Freedman, W. L., Aparicio, A., Bertelli, G., \&
Chiosi, C. 1999b, \aj, in press (astro-ph/9906121)
\reference{EG99} Grebel, E. K. 1999, in The Stellar Content of Local Group 
Galaxies, IAU Symposium 192, eds P. Whitelock \& R. Cannon, (San Francisco, 
ASP), p.\ 17
\reference{GG99} Grebel, E. K. \& Guhathakurta, P. 1999, \apj, 511, L101
\reference{GN90} Green, E. M., \& Norris, J. E. 1990, \apj, 353, L17
\reference{CG98} Grillmair, C. J., \etal 1998, \aj, 115, 144
\reference{GR97} Guarnieri, M. D., Renzini, A., \& Ortolani, S. 1997, \apj,
477, L21
\reference{WH82} Harris, W. E. 1982, \apjs, 50, 573
\reference{DH76} Hartwick, F. D. A. 1976, \apj, 209, 418
\reference{HK97} Hirashita, H., Kamaya, H., \& Mineshige, S. 1997, \mnras,
290, L33
\reference{H95A} Holtzman, J., \etal, 1995a, \pasp, 107, 156
\reference{H95B} Holtzman, J. A., Burrows, C. J., Casertano, S., Hester, J. J.,
Trauger, J. T., Watson, A. M., \& Worthey, G. 1995b, \pasp, 107, 1065 (H95)
\reference{HK99} Hurley-Keller, D., Mateo, M., \& Grebel, E. 1999, \apj, 523, 
L25
\reference{KK99}Karachentsev, I. D. \& Karachentseva, V. E. 1999, \aap, 
341, 355
\reference{KN93} K\"{o}nig, C. H. B., Nemec, J. M., Mould, J. R., \&
Fahlman, G. G. 1993, \aj, 106, 1819
\reference{RL74} Larson, R. B. 1974, \mnras, 166, 585
\reference{LM93} Lee, M. G., Freedman, W. L., \& Madore, B. F. 1993, \apj, 417,
553
\reference{SL47} Lee, S.-W. 1977a, \aaps, 27, 381
\reference{SL55} Lee, S.-W. 1977b, \aaps, 29, 1
\reference{YL89} Lee, Y.-W., Demarque, P., \& Zinn, R. 1994, \apj, 423, 248 
\reference{LB99} Lynden-Bell, D. 1999, in The Stellar Content of Local Group 
Galaxies, IAU Symposium 192, eds P. Whitelock \& R. Cannon, (San Francisco, 
ASP), p.\ 39
\reference{MF99} Mac Low, M.-M., \& Ferrara, A. 1999, \apj, 513, 142
\reference{SM99} Majewski, S. R., Siegel, M. H., Patterson, R. J., \&
Rood, R. T. 1999, \apj, 520, L33
\reference{MM98} Mateo, M. 1998, \araa, 36, 435
\reference{MR96} Mighell, K. J., \& Rich, R. M. 1996, \aj, 111, 777
\reference{MK90} Mould, J., \& Kristian, J. 1990, \apj, 354, 438
\reference{NF96} Norris, J. E., Freeman, K. C., \& Mighell, K. J. 1996,
\apj, 462, 241
\reference{OA96} Olszewski, E. W., Pryor, C., \& Armandroff, T. E. 1996,
\aj, 111, 750
\reference{RB85} Ratnatunga, K. U., \& Bahcall, J. N. 1985, \apjs, 59, 63
\reference{SF98} Schlegel, D. J., Finkbeiner, D. P., \& Davis, M. 1998, \apj,
500, 525
\reference{HS38} Shapley, H. 1938, Harvard College Obs.\ Bull., No.\ 908, 1
\reference{TS99} Smecker-Hane, T. A., Mandushev, G. I., Hesser, J. E.,
Stetson, P. B., Da Costa, G. S., \& Hatzidimitriou, D. 1999, in 
Spectrophotometric Dating of Stars and Galaxies, ASP. Conf.\ Ser., 
eds I. Hubeny, S. Heap \& R. Cornett, (San Francisco, ASP), in press
\reference{ST98} Stetson, P. B., 1998, \pasp, 110, 1448
\reference{PS99} Stetson, P. B., Bolte, M., Harris, W. E., Hesser, J. E.,
van den Bergh, S., VandenBerg, D. A., Bell, R. A., Johnson, J. A., Bond, H. E.,
Fullton, L. K., Fahlman, G. G., \& Richer, H. B. 1999, \aj, 117, 247
\reference{PS98} Stetson, P. B., Hesser, J. E., \& Smecker-Hane, T. A.
1998, \pasp, 110, 533
\reference{BT80} Tinsley, B. M. 1980, Fundamentals of Cosmic Physics, 5, 287
\reference{PV86} Vader, J. P. 1986, \apj, 305, 669
\reference{PV87} Vader, J. P. 1987, \apj, 317, 128
\reference{SB94} van den Bergh, S. 1994, \apj, 428, 617
\reference{AW92} Walker, A. R. 1992, \pasp, 104, 1063
\reference{AW94} Walker, A. R. 1994, \pasp, 106, 828
\reference{WH97} Whitmore, B., \& Heyer, I. 1997, Instrument Science Report 
WFPC2 97-08 (Baltimore: STScI) (WH97)
\reference{ZO89} Zaritsky, D., Olszewski, E. W., Schommer, R. A., 
Peterson, R. C., \& Aaronson, M. 1989, \apj, 345, 759
\reference{ZS93} Zaritsky, D., Smith, R., Frenk, C., \& White, S. D. M. 1993,
\apj, 405, 464
\reference{ZS94} Zaritsky, D., \& White, S. D. M. 1994, \apj, 435, 599
\reference{RZ78} Zinn, R. 1978, \apj, 225, 790
\reference{ZW84} Zinn, R., \& West, M. J. 1984, \apjs, 55, 45

\end{references}
\end{document}